\documentclass[aps,prb,twocolumn,floats,epsfig,groupaddress]{revtex4-1}
\usepackage{amssymb}
\usepackage{amsbsy}
\usepackage{amsmath}
\usepackage{xcolor}
\usepackage{bm}
\usepackage{epsfig}
\usepackage{color}
\usepackage{soul}
\usepackage{float}
\usepackage{braket}
\usepackage[colorlinks,linktocpage,bookmarks=false,citecolor=blue,linkcolor=red,urlcolor=blue]{hyperref}
\newcommand\mybar{\kern1pt\rule[-\dp\strutbox]{.8pt}{\baselineskip}\kern1pt}

\newcommand{\beq}{\begin{equation}}
\newcommand{\eeq}{\end{equation}}
\newcommand{\bea}{\begin{eqnarray}}
\newcommand{\eea}{\end{eqnarray}}

\newcommand{\SN}{S_{\rm N}}
\newcommand{\tr}{{\rm tr}}
\newcommand{\bracket}[3]{\langle #1 | #2 | #3 \rangle}

\usepackage{xcolor}

\begin{document}
\title{Resonance-induced growth of number entropy in strongly disordered systems}
\author{Roopayan Ghosh}
\author{Marko \v Znidari\v c}
\affiliation{Department of Physics, Faculty of Mathematics and Physics, University of Ljubljana, Jadranska 19, SI-1000 Ljubljana, Slovenia}

\begin{abstract}
We study the growth of the number entropy $\SN$ in one-dimensional number-conserving interacting systems with strong disorder, which are believed to display many-body localization. Recently a slow and small growth of $\SN$ has been numerically reported, which, if holding at asymptotically long times in the thermodynamic limit, would imply ergodicity and therefore the absence of true localization. By numerically studying $\SN$ in the disordered isotropic Heisenberg model we first reconfirm that, indeed, there is a small growth of $\SN$. However, we show that such growth is fully compatible with localization. To be specific, using a simple model that accounts for expected rare resonances we can analytically predict several main features of numerically obtained $\SN$: trivial initial growth at short times, a slow power-law growth at intermediate times, and the scaling of the saturation value of $\SN$ with the disorder strength. Because resonances crucially depend on individual disorder realizations, the growth of $\SN$ also heavily varies depending on the initial state, and therefore $\SN$ and von Neumann entropy can behave rather differently.
\end{abstract}
\maketitle

\section{Introduction}

Non-interacting disordered systems in one dimension are known to exhibit localization at any disorder strength -- the famous Anderson localization~\cite{PhysRev.109.1492}. Localization in interacting models~\cite{BASKO20061126,mirlin05} -- the so-called many-body localization (MBL) -- has been the topic of intense research in recent years~\cite{doi:10.1146/annurev-conmatphys-031214-014726,ALET2018498,RevModPhys.91.021001}. The current understanding, based on numerics and an almost rigorous proof for a particular model~\cite{Imbrie16}, is that there is a critical disorder strength beyond which the interacting systems do localize, which should then by definition of localization be reflected in lack of transport of conserved quantities such as energy and particles.

\begin{figure}
\includegraphics[scale=0.5]{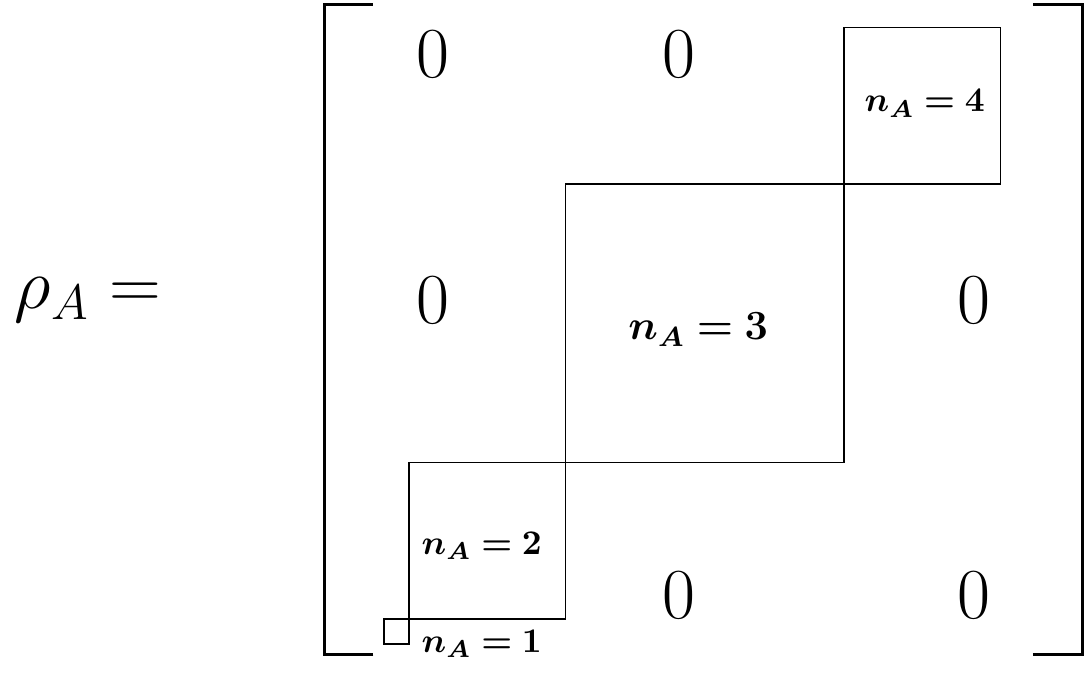}
\caption{A schematic diagram of the reduced density matrix $\rho_A$.}
\label{fig0}
\end{figure}

One of the surprising properties of the MBL phase is that despite the lack of transport one nevertheless does get the spreading of quantum information as indicated for instance by a logarithmically slow growth~\cite{PhysRevB.77.064426,PhysRevLett.109.017202} of bipartite von Neumann entropy $S(t)$, which one computes by dividing the system into two subsystems $A$ and $B$. Simply put, in a many-body localized system non-conserved degrees of freedom can still exhibit nontrivial dynamics which is reflected in the growth of $S(t)$ that in a finite system eventually saturates to a sub-ergodic volume law value~\cite{PhysRevLett.109.017202}. The logarithmic growth $S(t) \sim \ln{t}$ can be explained by the so-called local integrals of motion model~\cite{PhysRevB.90.174202,PhysRevLett.111.127201,serbyn14,anto15}, expected to hold for MBL systems, where it is due to dephasing from exponentially decaying coupling~\cite{PhysRevLett.110.260601,vosk14,kim14}. The entanglement entropy $S(t)$ is quite difficult to directly measure in experiments, however, in systems that conserve the total number of particles there is a much more accessible number entropy $\SN$. In such systems, starting with an initial state $\ket{\psi}$ that is an eigenstate of the total number operator, the reduced density operator $\rho_{\rm A}(t)$ will at all times have a block structure in the number (i.e., computational) basis (see Fig.~\ref{fig0}). Writing each block matrix containing $n_A$ particles as $p_{n_A} \rho_{n_A}$, with $\tr{\rho_{n_A}}=1$, we can write the von Neumann entropy $S(\rho_A)=-\tr{[\rho_A \ln{\rho_A}]}$ as a sum of two terms,
\begin{equation}
S=\SN+S_{\rm conf},
\label{eq:Ssplit}
\end{equation}
where $\SN=-\sum_{n_A} p_{n_A} \ln{p_{n_A}}$ is the number entropy, while $S_{\rm conf}=\sum_{n_A} p_{n_A} S(\rho_{n_A})$ is called the configurational entropy. The number entropy $\SN$ is much easier to measure because it requires only probabilities $p_{n_A}$ to find $n_A$ particles in the subsystem A. Experimental measurement of number entropy in Ref.~\onlinecite{doi:10.1126/science.aau0818} stimulated a flurry of numerical studies~\cite{PhysRevLett.124.243601,PhysRevB.103.024203,KIEFEREMMANOUILIDIS2021168481,kuba21} of $\SN$ in the Heisenberg model at strong disorder (considered to be deep in the MBL phase), fitting the numerical data by a slow growth of $\SN\sim \ln{\ln{t}}$  (this growth of $\SN$ is unrelated to a subleading~\cite{PhysRevB.97.214202} $\sim \ln{\ln{t}}$ growth of von Neumann entropy in a dephasing model of MBL) and arguing that this growth is unbounded which would challenge the established MBL phenomenology \cite{Note1}, specifically no particle transport in the localized phase. A subsequent study focusing on the steady state saturation value of $\SN$ on the other hand concluded~\cite{PhysRevB.102.100202} that there are no signs of ergodicity and that $\SN$ is compatible with MBL. 

In the present paper we resolve the issue by explaining that, in fact, all the above numerical observations are compatible and explainable by properties of the MBL phase. Taking into account rare resonances \cite{villalonga2020eigenstates,anushya20,PhysRevB.104.184203,david21,PhysRevB.92.104202,crowley2021partial} present in MBL systems we explain how this affects growth of $\SN$ in single disorder configurations and how these combine to give the slow growth after disorder averaging. With a simple resonance model we shall predict a slow power-law growth (instead of reported $\ln \ln t$) of $\SN(t)$ at intermediate timescales and an eventual saturation of $\SN$ to a nonergodic finite value whose scaling with disorder strength perfectly agrees with numerics. We remark that while our results do not resolve the recently hotly debated issue~\cite{jan20,dima21,EPL01,PhysRevLett.125.156601,kuba20,anushya20,anatoli21,david21} of whether a true MBL phase at all exists in the Heisenberg model, as our numerics is also limited to small systems of length $L \leq 18$ (which is not sufficient to resolve the question~\cite{EPL01}), it does show that all observable phenomenology of $\SN(t)$ at these sizes is fully compatible with MBL.

\section{Model}
\label{model}

We consider the spin-$1/2$ isotropic Heisenberg model with random magnetic field in the $z$ direction,
\begin{equation}
H=-J\sum_{i=1}^{L-1}\mathbf{S}_i \cdot \mathbf{S}_{i+1} +\sum_{i=1}^L h_i S_i^z
\end{equation}
where $\mathbf{S}$ denotes a vector of spin operators and $h_i$ are random variables uniformly drawn from $[-W,W]$, with $W$ being the disorder strength. Using the Jordan-Wigner transformation one can equivalently write this Hamiltonian in terms of spinless fermions,
\begin{equation}
H=-\frac{J}{2}\sum_j (c_j^{\dagger}c_{j+1}+ \text{h.c.})+\sum_j h_j c_j^{\dagger} c_j+J\sum_j n_jn_{j+1}.
\label{ham}
\end{equation}
where $c_j[c_j^{\dagger}]$ are the annihilation[creation] operators of the fermion and $n_j=c_j^{\dagger}c_j$.\cite{Note2} Without loss of generality we shall set $J=1$ throughout our work. Although one often refers to~\cite{palhuse,PhysRevB.91.081103} $W_{\rm c}\approx 3.7$ as the MBL transition point in this model, recent results~\cite{PhysRevX.7.021013,PhysRevLett.123.180601,elmer18,PhysRevB.101.035148,PhysRevResearch.2.032045,PhysRevLett.125.156601,PhysRevB.97.201105,PhysRevResearch.2.042033} indicate that there may be a significant shift toward larger disorders with increasing system size, perhaps to $W_{\rm c} \approx 5$ or even higher. To avoid any possible effects at $W_{\rm c}$, where one needs large systems~\cite{EPL01}, we will focus on $W\ge 10$ that is supposed to be in the MBL phase.

As the Hamiltonian conserves particle number we shall confine ourselves to the half-filled sector of the system. Unless otherwise mentioned, the initial states are chosen randomly from computational basis states, i.e., product states that are eigenstates of all $n_j$. We consider a half-half bipartition of the system throughout our work, and denote the two haves by $A$ and $B$. The reduced density operator and the von Neumann entropy $S$ are
\begin{equation}
\rho=\ket{\psi}\!\bra{\psi},\quad \rho_A=\tr_B[\rho],\quad S=-\tr[\rho_A \ln \rho_A]
\end{equation}
where $\ket{\psi}$ is the state of the full system (in numerical data we use natural logarithm). As we have outlined in Eq. \ref{eq:Ssplit}, in number-conserving systems one can split the von Neumann entropy into two terms; one is the number entropy $\SN$ which is just the Shannon entropy of the probability distribution $p_{n_A}$ to find $n_A$ particles in the subsystem A, and the other is configurational entropy that measures correlations between arrangements of particles of the subsystem and the environment. Physically speaking, the number entropy is a measure of entanglement generated by actual transport of particles across the boundary of the subsystem. Configurational entropy is more subtle. It takes into account all configurational correlations and thus unlike number entropy grows as the system undergoes dephasing to reach a steady state, even if there is no change in $n_A$.  Probabilities $p_{n_A}$ needed for the number entropy $\SN=-\sum_{n_A=0}^{L_A} p_{n_A}\ln p_{n_A}$, where $L_A$ is the length of subsystem A, are equal to the trace of the corresponding block of the block diagonal matrix $\rho_A$ (see Fig.~\ref{fig0}). Explicitly
\begin{equation}
p_{n_A}(t)=\sum_{\bracket{k}{N_A}{k}=n_A} \braket{k|\psi(t)}\braket{\psi(t)|k},
\end{equation}
where $N_A=\sum_{j=1}^{L_A} n_j$. In the MBL phase, due to localization, there is no particle transport at long times. Therefore $\SN$ should saturate to a non-ergodic value.

\begin{figure*}
\includegraphics[width=0.7 \columnwidth]{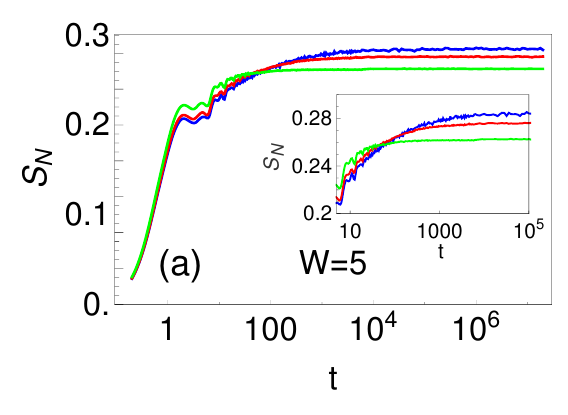}
\includegraphics[width=0.7 \columnwidth]{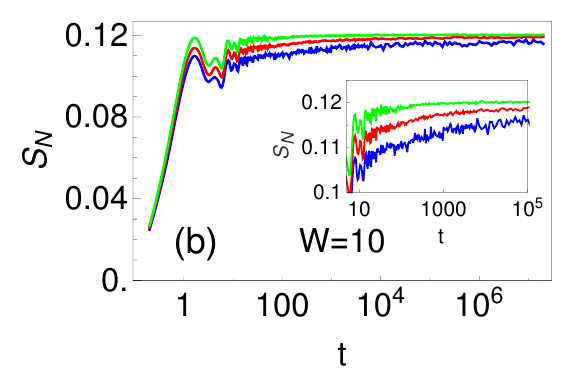}\\
\includegraphics[width=0.7 \columnwidth]{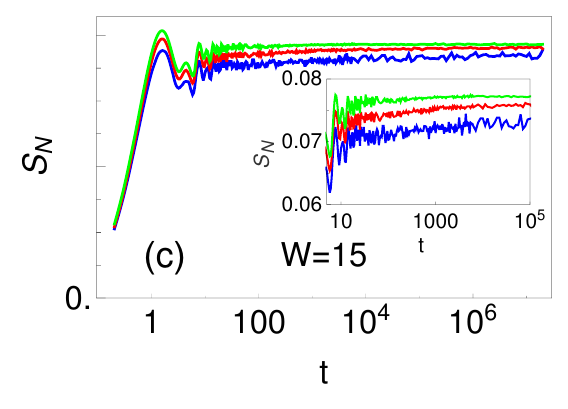}
\includegraphics[width=0.7 \columnwidth]{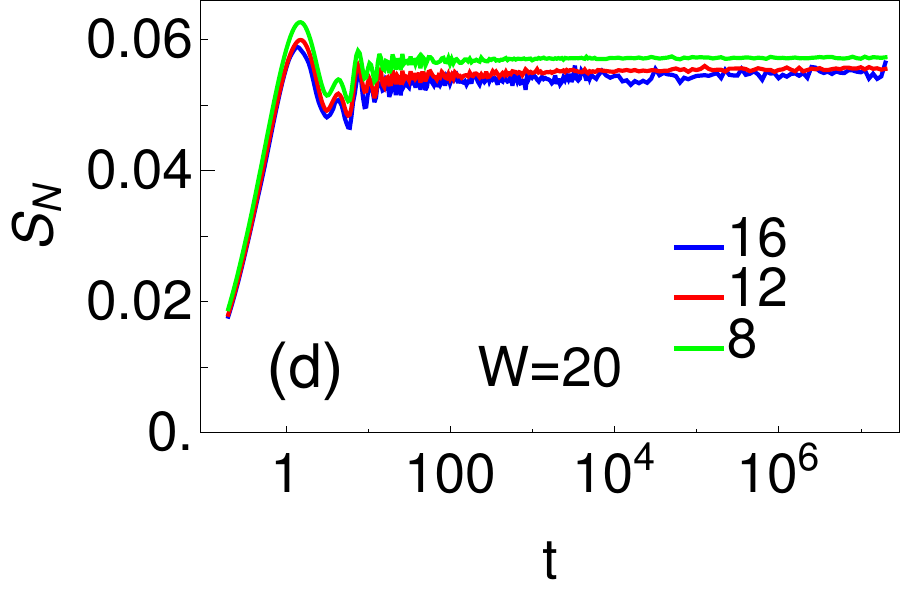}
\caption{Plots of $\SN(t)$ for different system sizes ($L=8,12,16$, with green, red, and blue, respectively) for the isotropic Heisenberg chain (Eq.~\ref{ham}), with a random magnetic field of strength $W$. The insets zoom in on the region of small growth at long times.}
\label{fig1}
\end{figure*}
In Fig. \ref{fig1} we have plotted mean $\SN$ for different disorder strengths $W$. Data for $L=8,12,16$ have been averaged over $5 \times 10^5$, $10^5$ and $10^4$ configurations, respectively (a configuration means an independent disorder realization and an independent half-filling product initial state). We notice that there is a clear but small growth of $\SN$ for all the system sizes.

Before jumping to speculation that this indicates ergodicity and therefore the absence of MBL a few observations are in order. (i) Growth is in all cases rather small. (ii) The case of $W=5$, where it is the largest, is close to $W_{\rm c}$ (or even in the ergodic phase). (iii) While one can fit $\SN \sim \ln{\ln{t}}$ to data~\cite{PhysRevLett.124.243601,PhysRevB.103.024203,KIEFEREMMANOUILIDIS2021168481} the agreement is limited to a rather tiny window of $\SN$\cite{Note1}. Many other functions could be fitted with similar significance; in fact, as we will see, a power-law predicted theoretically gives a better fit over wider range. (iv) Saturation values of $\SN$ are in all cases small and far from being ergodic. For example, a random half-filled state to which one would converge at long times in an ergodic system has $\SN \asymp \frac{1}{2}\ln{(e \pi L/8)}$, i.e., $\SN \approx 1.45$ at $L=16$ and $\SN \approx 1.14$ at $L=8$. Therefore, saturation values are already at $W=10$ more than $10$ times smaller than the ergodic ones. (v) The saturation value of $\SN$ shows no significant increase with $L$ , if anything, at large $W$ it decreases with $L$.

In the following sections we will analyze the cause of the growth and show it is compatible with localization. First in Sec. \ref{resonant}, we will talk about special resonant configurations where one sees a sharp rise in $\SN$ at a particular time, and which can be held accountable for a small increase visible in the mean $\SN$. Then in Sec. \ref{steady} we shall investigate the steady state behaviour of the system, and show how several observed features can be well explained by a two-state resonant model. This is followed by Sec. \ref{short}, where we shall discuss the short time behaviour of the system and how choice of initial states affects the growth of mean $\SN$. After that in Sec. \ref{inter} we shall show how one can kill the growth of $\SN$ by filtering out a few special resonant configurations. We will also discuss the nature of growth of $\SN$ at intermediate times in this section. 

\section{Resonant configurations}
\label{resonant}

\begin{figure}
\includegraphics[width=0.75 \columnwidth]{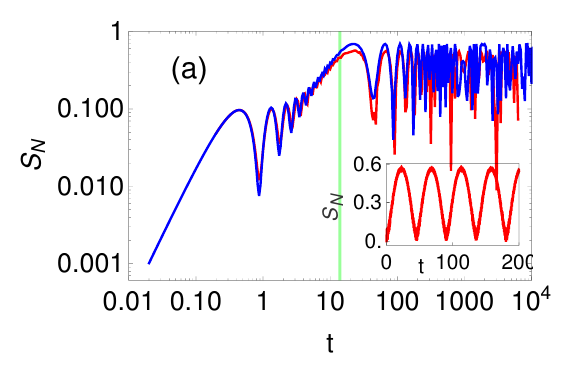}
\includegraphics[width=0.75 \columnwidth]{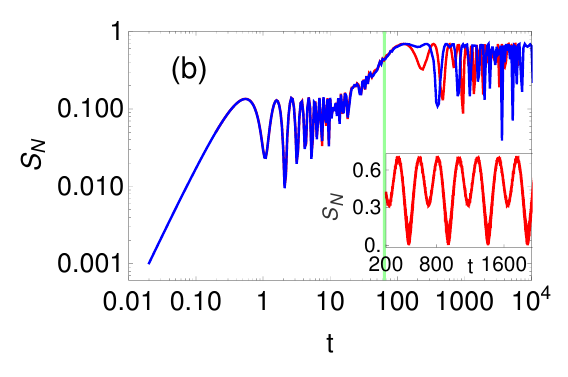}
\caption{Plots of $\SN(t)$ for two resonant configurations starting from the domain wall initial state at $W=15$ and $L=12$. The red lines are from exact numerics, the blue lines are the result we get from the $m$-state problem [with $m=3$ for (a) and $m=5$ for (b)]. The green line denotes the time $1/(E_1-E_2)$ at which the resonance is resolved and $\SN$ exhibits a jump. The insets show the numerically obtained $\SN$ vs $t$ in the main plot for a shorter timespan on a linear scale, to make the Rabi-like oscillations visible.}
\label{fig1a}
\end{figure}

In Fig.~\ref{fig1a} we plot $\SN$ for two specific configurations where one can see an interesting behaviour of $\SN$: upto potentially long time $\SN$ exhibits a plateau, like it would already saturate to its long-time value, but then jumps to new larger saturation value. Such configurations exhibiting a plateau with a jump are rare (at large $W$), with the time of the jump very much depending on the configuration. As we shall explain in the following, it is resonances that are responsible for such jumps and they will form the core of our theory of $\SN$ growth. For instance, after averaging over different configurations with different jump times they will result in the slow growth of $\SN$ visible in Fig.~\ref{fig1}.

To understand why these resonances occur we first consider the simpler case of very large $W$. In such a case, one would expect the eigenfunctions to be localized strongly at the corresponding computational states, and eigenvalues equal to the diagonal elements of the Hamiltonian. The resonances occur when two computational states which are very close to each other in unperturbed energy, hybridize with each other. In such a scenario, if one starts from any one of the computational states, it mixes strongly with the other over time. Crucially for $\SN$, if these computational states possess a different particle number for subsystem $A$, then this hybridization shows up as a `jump' in $\SN$. The time of the jump is in the order of $1/(E^0_1-E^0_2)$, where $E^0_1=\bra{1}H\ket{1}$ and $E^0_2=\bra{2}H\ket{2}$, $\ket{1}$ and $\ket{2}$ denote the computational states involved in the hybridization.
 
 As we reduce $W$ the localization becomes weaker and one needs to examine the true eigenenergies instead of just the diagonal matrix elements. Additionally, the resonant eigenvectors also have significant overlaps with more than one computational state. In this situation, hybridization can occur when the eigenvectors associated with eigenenergies close to each other have a large overlap with two computational states having different $n_A$. Then if the initial state is one of the two computational states involved, the time of the jump can be approximately given by $1/(E_1-E_2)$, where $E_1$ and $E_2$ are eigenenergies of the relevant states.
 
We can show an easy demonstration of the above in some special cases, where one can approximate such a situation via $m \times m$ block of the Hamiltonian, when $m$ is small enough and is similar to the range of hybridization. The range of hybridization is given by the number of `hops' required to go from the initial state to the final. In Figs. \ref{fig1a}(a) and (b) we plot $\SN$ vs $t$ for two distinct disorder realizations with the same initial state $\ket{1}$. The states involved in hybridization in Fig. \ref{fig1a}(a) are $\ket{1}=\ket{111111;000000}$ and $\ket{2}=\ket{111110;010000}$ which have a range of hybridization $2$ (a semicolon is used to show the separation between systems $A$ and $B$). The disorder configuration in this scenario was such that one had a resonance. Specifically $h_6=5.08$ and $h_8=6.07$ were resonant so that the 0th-order energies of states $\ket{1}$ and $\ket{2}$, in which a particle jumped from the 6th to 8th site, were resonant, $|h_6-h_8+1| \ll 1$. The analytical curve in blue is found from taking a 3 -state model with these two states and the intermediate $\ket{3}=\ket{111110;100000}$, and calculating $\SN$ for evolution on this $3$-dimensional subspace. It shows a very good match with the numerical plot in red. Since the two states involved differ in number of fermions in subsystem $A$, there would be an increase in $\SN$ when one starts from any one such state initially in the Hamiltonian.  In the plot we also show the timescale of such a jump indicated by the green line. Since $\SN$ grows with time as $\sim \sin^2((E_1-E_2)t)$, where $E_1$ and $E_2$ are eigenenergies of the eigenstates involved in hybridization, the rise can be seen around time $t\sim 1/(E_1-E_2)$.

In Fig. \ref{fig1a}(b) we start from the same state but look at a resonance with range of $3$; i.e., the final state is given by $\ket{111110;001000}$. This corresponds to a longer range hybridization. Here we need the two intermediate states and another state $\ket{111110;000100}$, i.e. a total of 5 states to obtain $\SN$ accurate to what we get for the full system. This is because in the configuration chosen $\ket{5}=\ket{111110;000100}$ is resonant to $\ket{3}=\ket{111110;100000}$ alongside  $\ket{1}=\ket{111111;000000}$ being resonant to $\ket{4}=\ket{111110 ;001000}$, and this double resonance allows a strong hybridization between $\ket{1}$ and $\ket{4}$ even when they are separated by 3 lattice hops in a strong disorder regime. In this case the 0th-order resonance condition reads $|h_6-h_9+1| \sim |h_7-h_{10}| \ll 1$, which triggered the double resonance.

While the domain wall state was chosen for easy demonstration of the phenomenon, behavior of other generic initial states at strong disorder can also be approximated by such an effective low-dimensional model, resulting in qualitatively same observation. When we calculate mean $\SN$ we average over many configurations showing the resonant behaviour in different timescales. This shows up as a slow growth in mean $\SN$. For growth at late times one requires presence of resonant cases at times $t \gg 1$, and this is possible when two resonant states have $E_1-E_2 \ll 1$. We have noticed that cases of long range resonances are more abundant in the disorder strength range $W\sim5-10$ in agreement with what has also been reported in Ref.~\onlinecite{david21}. Consequently, from Fig. \ref{fig1} we can also clearly see that the increase in $\SN$ is more visible in this region, where larger system sizes clearly show larger values of $\SN$. At disorder strengths larger than this such strong resonances are very rare and the growth is stifled. These longer-range resonances cause $\SN$ to rise at later-times and thus they are the ones to drive the later time growths. The reason is that long-range resonances are driven by comparatively smaller eigenenergy differences. Since the approximate time of growth is inversely proportional to it, the rise occurs at later times, a feature seen when one compares Figs. \ref{fig1a} (a) and (b).

Additionally, since only a few states are involved in the hybridization and contribute to growth of $\SN$ one can expect Rabi like oscillation to be visible in its growth with time. This feature is shown in the insets of Fig. \ref{fig1a}. It is interesting to note that, the more the number of states involved in the hybridization, the more complicated the oscillations are, which is what one expects.

 The initial state in the demonstration was chosen such that it had a very low connectivity locally in the Hilbert space. This allowed us to approximate the effective low-dimensional model directly from a block in the Hamiltonian. However, if one starts from a general computational initial state and not so strong disorders, one would need to take into account a large number of intermediate states between the two resonant states to obtain a reasonable approximation of $\SN$. The strength of hybridization also depends strongly on the intermediate states involved. Due to the energy difference between the states being very small, a perturbative approach to calculate the effective Hamiltonian to show the hybridization is very difficult and one needs to go to very high orders of perturbation to obtain a sensible result. In such more complicated single-resonance cases, the technique described by Ref.~\onlinecite{david21} to find the effective $2$-state Hamiltonian between the resonant states is a better approach to visualize it semi-analytically. This is discussed in Appendix \ref{effham}.

In Sec. \ref{inter} we shall discuss more on the role of resonances to increase mean $\SN$ and predict the law of growth to follow a power law, using statistics of the resonant configurations. This will be shown to agree very well with the numerical results.

\section{steady states}
\label{steady}
In this section we shall describe the statistics of the long-time saturation values of $\SN$ (steady state values) for different disorder strengths and system sizes. We shall discuss the power law decrease of mean and median steady state $\SN$ with disorder strength, which we obtain numerically and provide an analytical basis for the same. We will also show how steady state $\SN$ decreases with system size for large $W$, confirming lack of ergodicity.

\subsection{Statistics of steady state quantities}
There are two comparable quantities one usually looks at while studying long-time steady state properties of a system. One is the long-time average of $\SN$, denoted by $\tilde{S}_N$, which we calculate\cite{Note3} by averaging the data between $t \sim 10^6-10^7$. The other quantity, which we denote by $\bar{S}_N$, is calculated by first calculating $\bar{p}_{n_A}$, the steady state probability of having $n_A$ particles in the subsystem $A$, and then calculating $\bar{S}_N=-\sum_{n_A=0}^{L_A} \bar{p}_{n_A} \ln \bar{p}_{n_A}$. We know from  Jensen's inequality that $\tilde{S}_N \le \bar{S}_N$ for any set of parameters since $\SN$ is a concave function of $p_{n_A}$. Hence, in what follows we shall concentrate on the properties of $\bar{S}_N$. Some results for $\tilde{S}_N$ are presented in Appendix \ref{appB1}.

\begin{figure}
\includegraphics[width=0.98 \columnwidth]{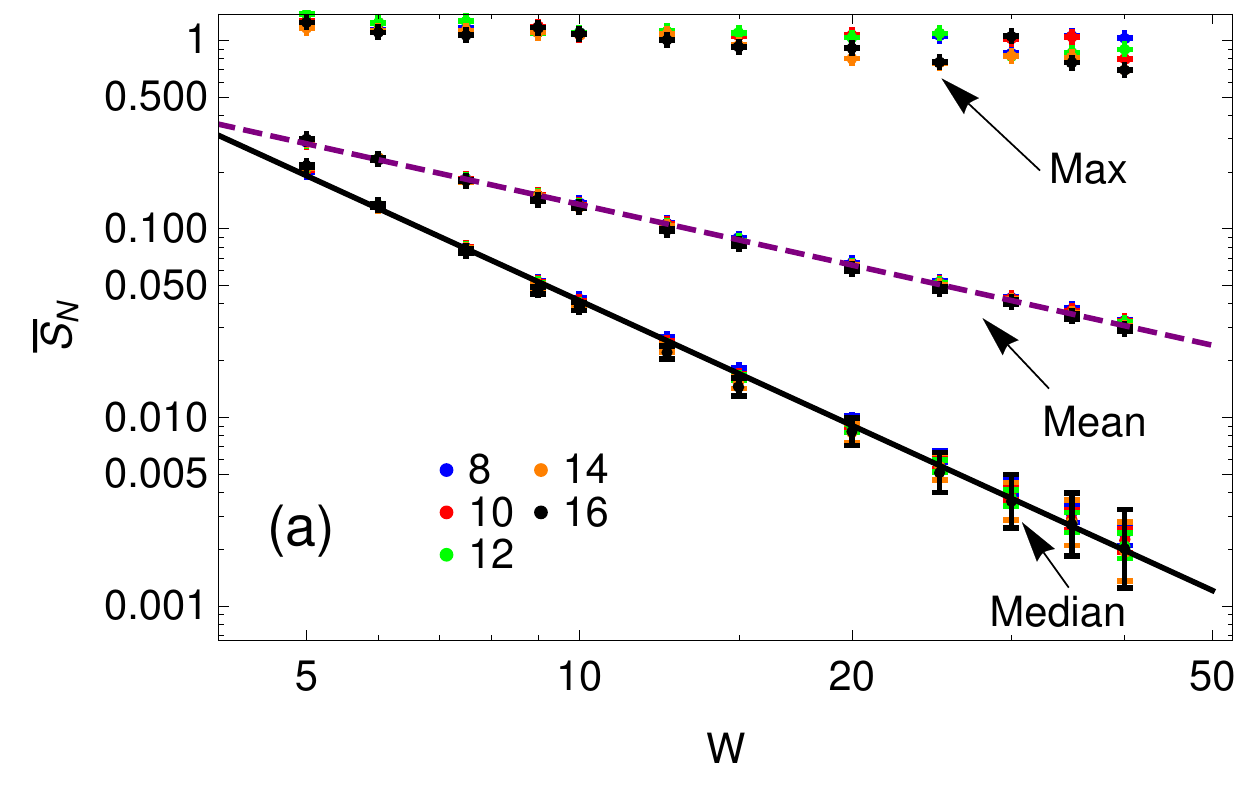}
\hspace{0.1 in}\includegraphics[width=0.9 \columnwidth]{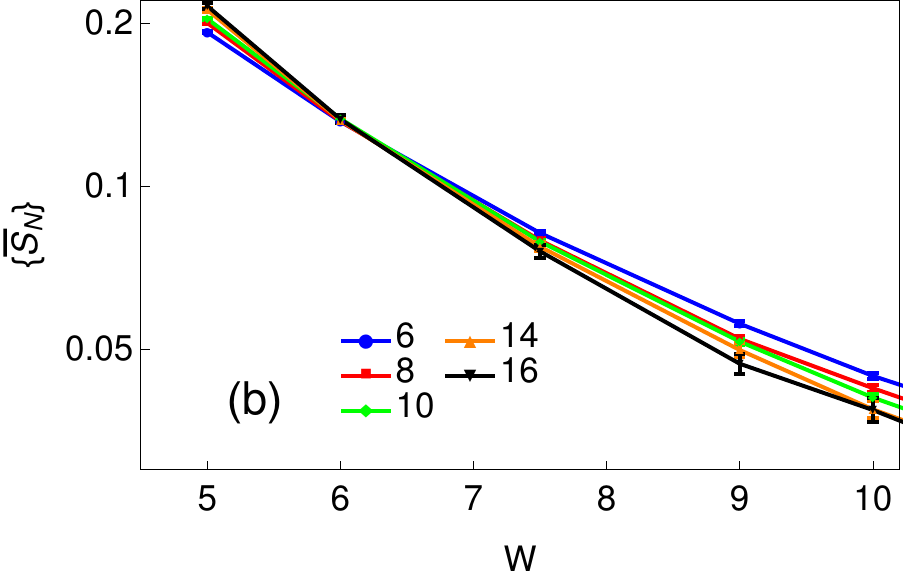}
\caption{ Plot showing long-time saturation values (steady state) of $\SN$ ($\bar{S}_N$) for different $W$ and $L$ (denoted by different colours). (a) Maximum, mean, and median of the distribution. The dotted line shows the fit to $W^{-1.07}$ and the solid line shows fit to $W^{-2.2}$. (b) Zoomed-in plot of the median ($\{\bar{S}_N\}$) from the top plot.}
\label{fig9}
\end{figure}
In Fig. \ref{fig9} we plot the maximum, mean and median of $\bar{S}_N$ for different system sizes and disorder strengths. We use $5 \times 10^5, 5 \times 10^5,10^5,10^5$ and $10^4$ configurations for $L=8,10,12,14$ and $16$ respectively. The maximum value obtained from the distribution remains more or less constant with increase in disorder and system size and is $\sim 1$. Considering how both mean and median deviate  from this value one can say that the maxima occur via the rare resonant cases which we talked about in the last section. Clearly they become rarer as $W$ increases and mean and median show a power law decrease with increasing $W$.  Furthermore we see that the mean is significantly larger than the median and falls much slower with $W$ than the median does.  This is generally expected for a distribution which is skewed toward $0$. We also see that the change of both mean and median with $W$ can be fitted by a power law with exponents very close to $-1$ and $-2$ respectively. Comparable features are also seen in Anderson localized systems, see Appendix \ref{anderson} for details.

We also want to point out that additionally, the trend of mean and median $\bar{S}_N$ with increasing $L$ is non-monotonic with increasing $W$. Figure \ref{fig9}(b) shows the median where the feature is prominent. Initially, for around $W=5$ it seems larger system sizes have a slightly larger value of both mean and median $\bar{S}_N$. But there is a reversal of this trend around $W\sim 6$, where beyond that, the larger system sizes show a lower mean and median $\bar{S}_N$ than smaller ones. This suggests that the exponents of the power law fits show a slight drift in value with increase in system size (see Appendix \ref{appB1}). However, it is to be noted that this behaviour with increasing $L$ supports the notion of lack of particle transport since distributions are further skewed towards $\SN=0$ for larger system size. This feature was also reported in Refs.~\onlinecite{PhysRevB.102.100202,PhysRevB.103.024203}. The trend while a bit surprising is actually caused due to the choice of initial states. This is discussed in the next section and in Appendix \ref{appB1}.

In Fig. \ref{fig10} we show the disorder averaged distribution of the change in number of particles in subsystem $A$ from what was present initially, to at the steady state, for different disorder strengths and $L=12$.  Clearly the disorder averaged value of the quantity $\delta n=|n-n_I|$, where $n_I$ is the number of particles initially present at the subsystem, fall off at least exponentially fast with distance. In fact for $W\ge 10$ clearly the fall is even faster than exponential. This observation further confirms lack of particle transport in the system as a whole, as if the particles indeed spread throughout the system, $p(\delta n)$ should not show such a fast decrease.  It also suggests that if any distribution which falls off slower than the exponential is indeed found in intermediate times, it is transient.

 Henceforth, unless otherwise mentioned, we shall use $\bar{S}_N$ to denote the ensemble mean of the steady state number entropy and $\{\bar{S}_N\}$ to denote median.

 While finding an analytical explanation for the power law decrease of $\bar{S}_N$ with $W$ from the many-body setup is difficult, one can still extract the trends at $W \gg J$ from a two-state model which we describe below. 

\subsection{Analysis via two-state model}
At large enough disorders the localization length is $\zeta<1$, and the distribution of change of particles in system $A$, $\delta n$ is very narrow around $\delta n=0$. This justifies an effective two-state model we are going to use. We group the probabilities into two, $\bar{p}$ and $1-\bar{p}$ where $\bar{p}$ denotes the steady state probability of the subsystem having different number of particles than in the initial state, $\ket{1}$. Due to the skewness of the distribution, the value of $\bar{p}$ is very close to $0$ for most cases, and hence one can approximate the statistics of $\bar{S}_N$ by the statistics of $\bar{p}$. 
\begin{figure}
\includegraphics[width=0.9 \columnwidth]{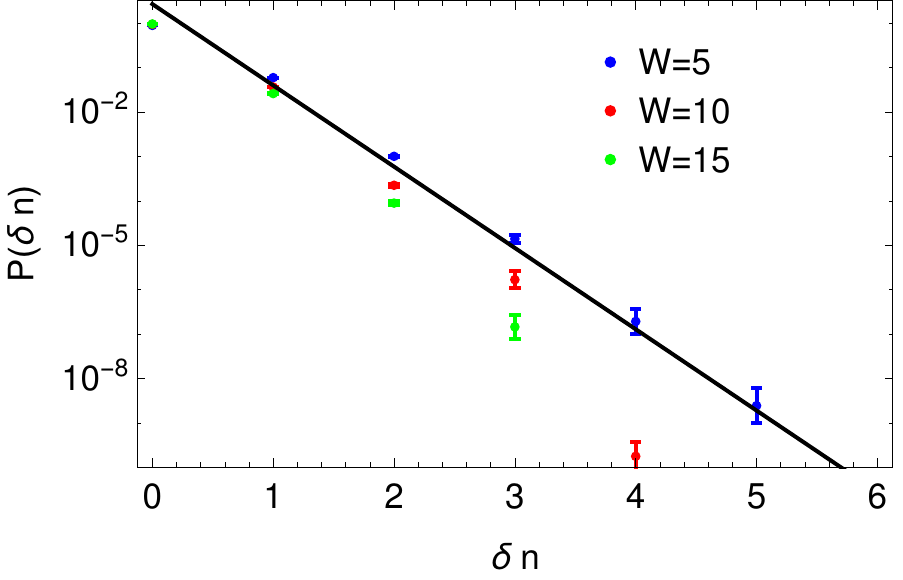}
\caption{Disorder-averaged distribution of probability of subsystem $A$ having $\delta n$ change in number of particles from its initial state, when it reaches long-time steady state. That is, $\delta n=|n-n_I|$, where $n_I$ is initial number of particles present in subsystem $A$, averaged over $10^6$ configurations for $L=12$. The black line denotes fit to $\exp( -\delta n/0.24)$.}
\label{fig10}
\end{figure}

We note that such simple two-dimensional description of resonances has appeared before in the context of MBL, for instance in describing hybridization probability~\cite{villalonga2020eigenstates}, contribution to 2-point level correlations due to resonances~\cite{PhysRevB.104.184203}, thermalization of a 2-level system coupled to a bath~\cite{crowley2021partial}, or to describe critical behavior observed in finite systems~\cite{anushya20,david21}. We consider a two state model whose Hamiltonian is given by, 
\begin{equation}
\begin{pmatrix}
E_1^0 &-j \\
-j & E_2^0
\end{pmatrix}
\label{twostate}
\end{equation} 
For nearest neighbour resonant states, this is a $2 \times 2$ block of the original Hamiltonian, which was also the low-dimensional model used in Sec. \ref{resonant}. Then $j=J/2$. However for more complicated and longer range resonances, we require an effective two dimensional model to describe them (for instance, see Appendix \ref{effham}). Then $E_1^0$ and $E_2^0$ will deviate from the corresponding diagonal elements of the Hamiltonian and $j\le J/2$. But, as we shall see in what follows it is not necessary to exactly formulate the effective low dimensional model to extract the power laws. The wavefunction evolves with time under action of this Hamiltonian as  $\ket{\psi(t)}=\sqrt{1-[d(t)]^2}\ket{1}+d(t)\ket{2}$ where
\begin{equation}
d(t)= A \sin(\Omega t/2).
\label{dt}
\end{equation}
The amplitude of oscillation is given by
\begin{equation}
A=\frac{2 j}{\Omega},
\label{amp}
\end{equation}
and the frequency is given by 
\begin{equation}
\Omega=\sqrt{(E_1^0-E_2^0)^2+4j^2}.
\label{freq}
\end{equation}
The simplification in analysis arises from the fact that this two state model simulates the jump of one particle from one site to another and we require only $E_1^0-E_2^0$ in our expressions. Hence, one can just take into account the on-site random numbers involved in the process, plus the interaction if necessary.  For this model, one can write down the evolution of $\SN$ with time as,
\begin{equation}
\SN(t)=\mathcal{H}(p(t)),
\label{SNexp}
\end{equation}
where $\mathcal{H}(x)=-x \ln x-(1-x) \ln (1-x)$ and $p(t)=[d(t)]^2$. The long time average of $\SN$ denoted by $\tilde{S}_N$ is equal to $\bar{\mathcal{H}}(p)$. However as discussed before, we are going to focus on the quantity $\bar{S}_N=\mathcal{H}(\bar{p})$, the steady state value, which due to concavity of the function is larger than $\bar{\mathcal{H}}(p)$. $\bar{p}$ is the long time average of $p(t)$. Due to skewness of the distribution, one can approximate $\bar{S}_N$ to be a linear function of $\bar{p}$ and hence expect $\bar{S}_N$ to show a similar behaviour to that of $\bar{p}$. We can hence focus on calculating statistics of $\bar{p}$ which is easier to compute and consequently understand the behaviour of $\bar{S}_N$.
From Eqs. \ref {dt}, \ref{amp} and \ref{freq}, we have,
\begin{equation}
\bar{p}=\frac{2j^2}{(E_1^0-E_2^0)^2+4j^2}.
\label{psteady}
\end{equation}  We will analyze the mean $\langle\bar{p}\rangle$ and median $\{\bar{p}\}$ of this quantity. 
We work in the large disorder region where we have $W \gg j$. From an analysis outlined in Appendix \ref{appB}, where using the fact that $E_1^0$ and $E_2^0$ are two random real numbers taken from a uniform distribution, one can find the distribution of $\bar{p}$ using two successive variable transformations, and hence see that for $j/W\ll 1$,
\begin{align}
\langle\bar{p}\rangle&=\frac{\pi j}{2W} +O(1/W^2) \nonumber \\
 \{\bar{p}\}&=(\frac{j}{W})^2(3+2\sqrt{2})+O(1/W^4), 
 \label{powerlaw}
\end{align}
Due to the presence of the higher order correction terms, one expects the observed powers to be slightly larger than $1$ and $2$ respectively for the mean and median\cite{Note4}.  Different scaling of the mean and median has been observed also in the distribution of eigenstate entanglement of a two-level spin coupled to a bath and which can also be described by a 2-level resonant model~\cite{crowley2021partial} resulting in a bi-modal distribution, similarly to our case.

\begin{figure}
\includegraphics[width=0.7 \columnwidth]{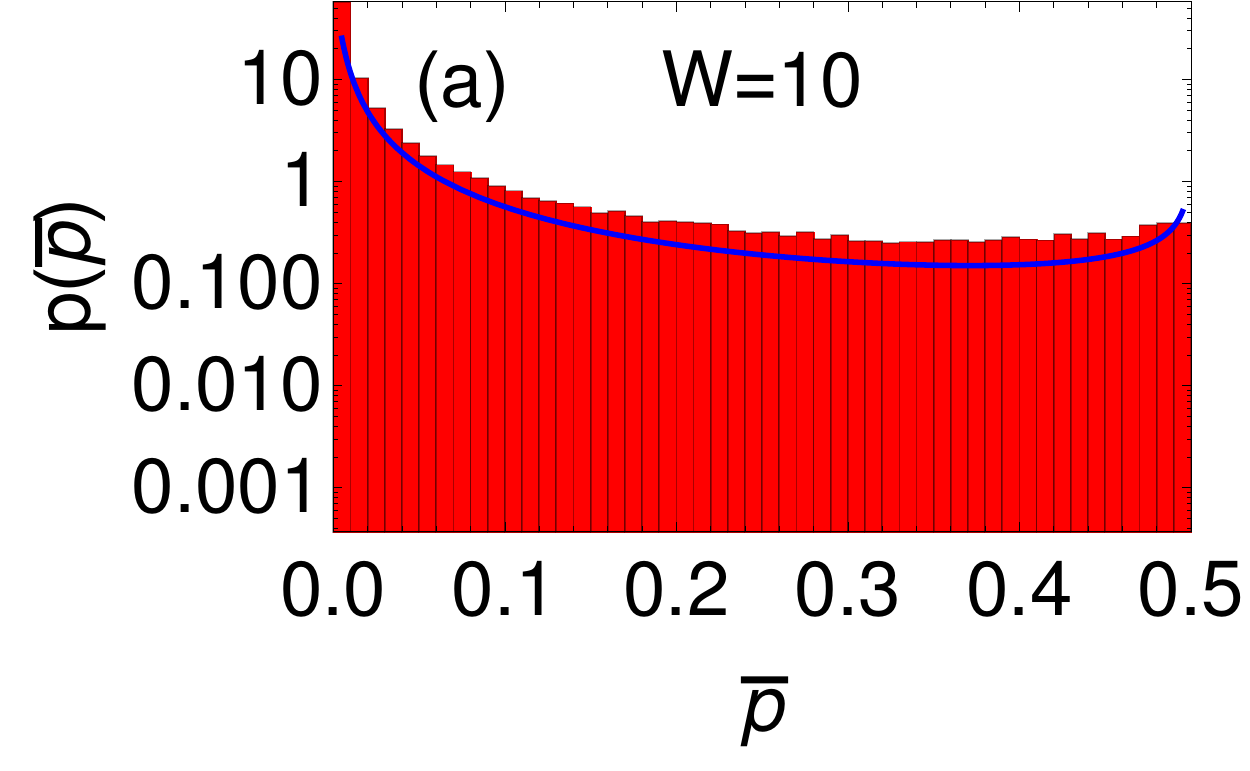}
\includegraphics[width=0.7 \columnwidth]{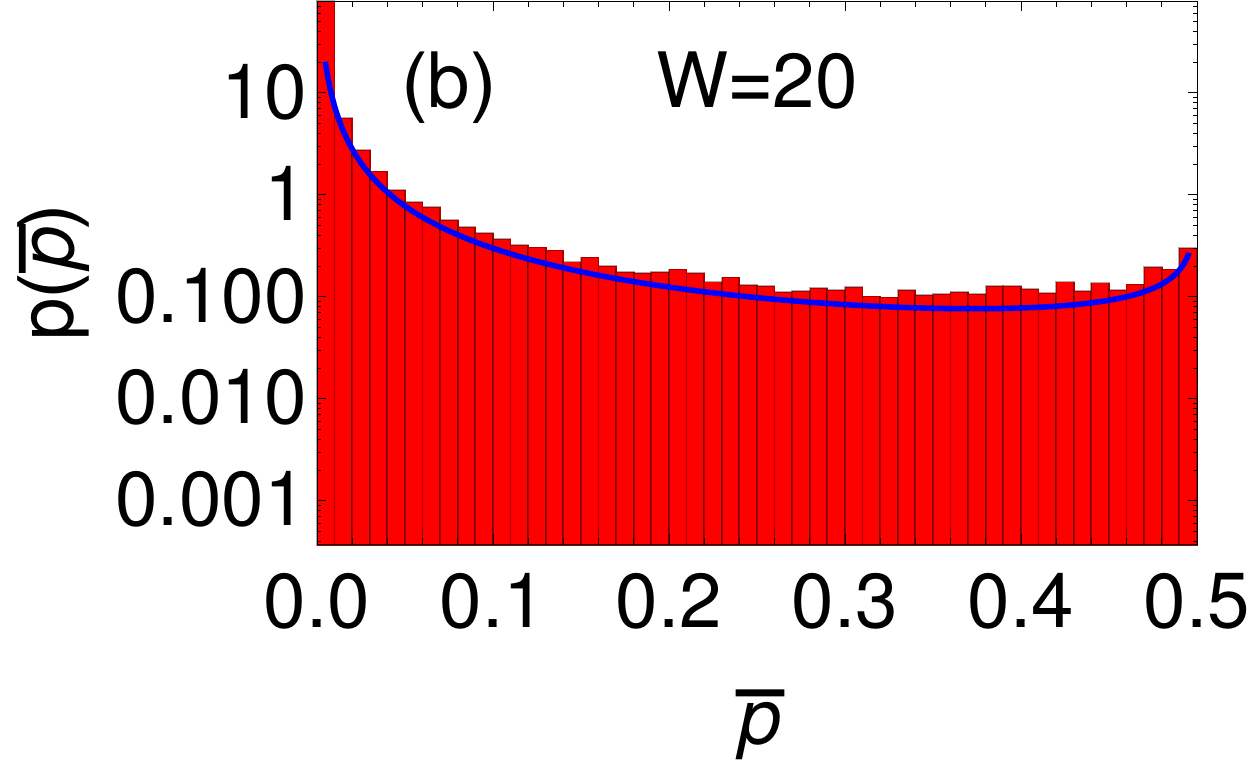}
\caption{Plot showing probability distribution of long time steady state probability $\bar{p}$ for $L=12$. The blue line is the distribution obtained from Eq. \ref{distrib} in App. \ref{appB} where we use the approximation $j=J/2=1/2$. The red bars are from exact numerics.}
\label{fig8b}
\end{figure}

These expressions show how the mean and median of $\bar{p}$ follow power laws with exponents close to what was shown in Fig. \ref{fig9}. Additionally in Fig. \ref{fig8b} we compare the probability distribution of $\bar{p}$ for disorder strength $W=10$ and $W=20$, for a system size of $L=12$ obtained from exact numerics (red histograms) with the probability distribution of $\bar{p}$ (Eq. \ref{distrib} in Appendix \ref{appB}) used to compute the quantities in Eq. \ref{powerlaw} (blue line). We see a very good agreement between the two, which gets better with increasing $W$ as expected. This strengthens the validity of our two state model, as not only the mean and median but the entire distribution can be approximated very well by the model. Some more finer aspects of the quantities in focus in this section are discussed in Appendix \ref{appB1}.

To summarize, analysis of steady state $\SN$ points toward the absence of any ergodicity at these disorder strengths in the model. We have shown how mean and median steady state $\SN$ follow power law decay with increasing $W$, with analytical support from a two-state model which almost correctly predicts the value of the exponent. Furthermore the probability of change in particles in the subsystem at the steady state falls at least exponentially with the change in the number. Also, for comparatively larger disorder strengths mean and median steady state $\SN$ shows a trend of decrease with system size rather than an increase. All of these features point to localization and we see no signature of any long range transport in the system.

 \section{Initial states and short time behaviour}
\label{short}
In this section we shall discuss the short time behaviour of $\SN$ which is approximately until $t\sim 1$.  In this time scale very few states are involved in the dynamics and the behaviour of $\SN$ can again be approximated by two-state models which we shall discuss in this section. Alongside this we shall also discuss how the behaviour of $\SN$ depends on the choice of initial states. The choice of initial states play a major role in the growth of $\SN$ at all timescales, but this effect is easiest to understand at such low timescales.
\subsection{Growth at small timescale}
\begin{figure}
\includegraphics[width=0.8 \columnwidth]{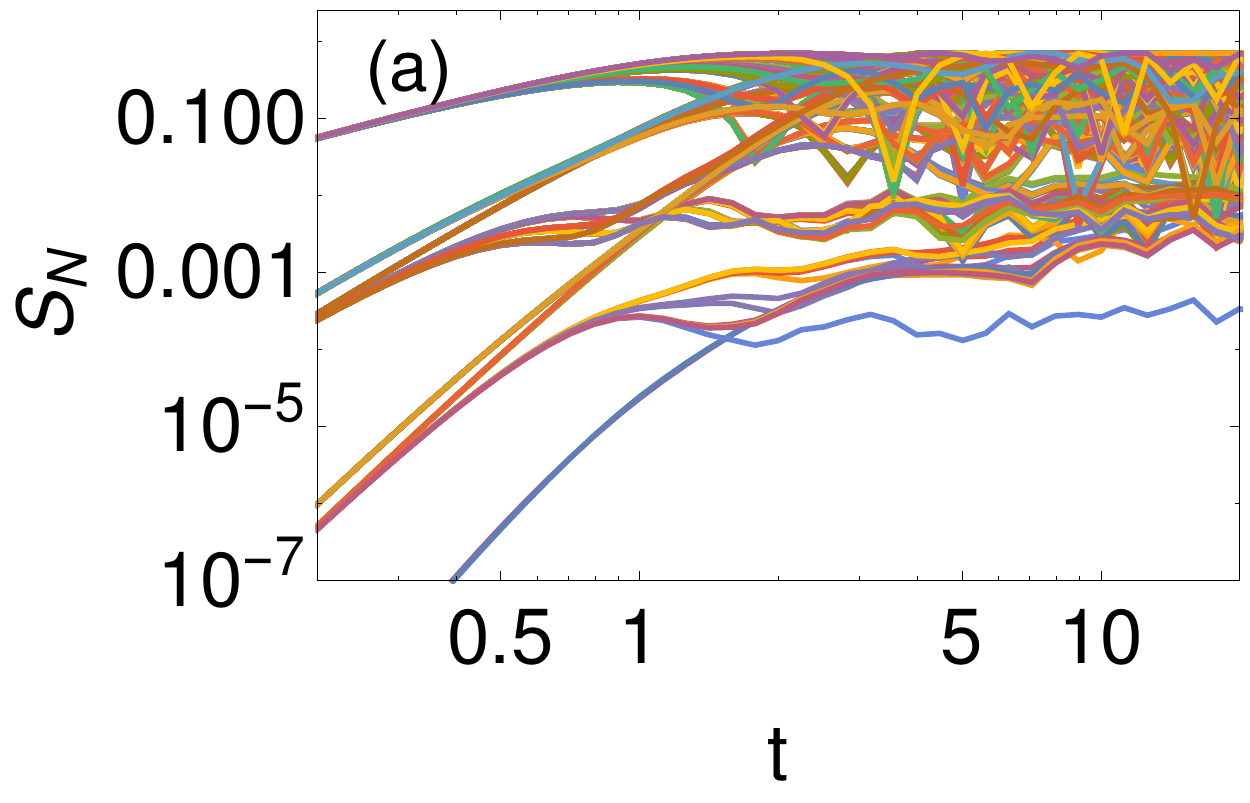}
\includegraphics[width=0.8 \columnwidth]{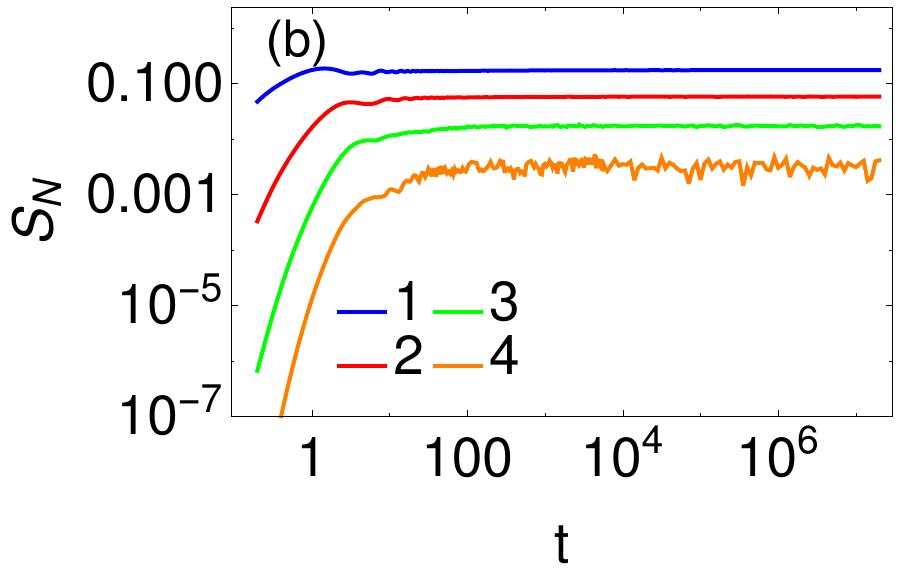}
\caption{Plot showing different behaviour of $\SN(t)$ for different groups of initial states. (a) Plots of $\SN(t)$ for a randomly chosen disorder realization for $W=10$ and $L=12$ for all allowed computational initial states. (b)Plot of configuration averaged $\SN(t)$ for groupings of different $q$ states taken from an ensemble of $10^5$ configuration of disorder and initial states for $L=12$ and $W=10$.}
\label{fig7}
\end{figure}

First in Fig. \ref{fig7}(a), we take one disorder realization and all possible computational initial states for $L=12$ and plot the growth in $\SN$ on the short timescales. One sees a clear grouping of initial states during the timescales $t <1$. Initially, we see that there are four groups of initial states each with a different power law growth of $\SN$.  We label each group by an index $q$, where $q=1 \hdots 4$. These groupings are done based on the number of consecutive $0$'s or $1$'s present around the subsystem cut. Let us label the sites in $A$ and $B$ by their distance from the cut. If one starts the labeling from $1$, $q$ denotes the smallest label where either the fermion occupancies in $A$ and $B$ do not match or have a different value from the occupancy at smaller labels.

To illustrate, let us describe in details the situation for $L=12$. The $q=1$ group consists of two kinds of states $\ket{\hdots0 ;1\hdots}$ and $\ket{\hdots1 ;0\hdots}$. Then the second group denoted by $q=2$  comprises four kinds of states $\ket{\hdots10 ;0\hdots}$,
 $\ket{\hdots0 ;01\hdots}$ and their spin flipped versions. The third one 
 $q=3$ has another four kinds of states $\ket{\hdots 100 ;00\hdots}$, $\ket{\hdots00 ;001\hdots}$ and their spin flipped versions, and finally in the last group labeled by $q=4$ one has the remaining two states viz. $\ket{000111 ;111000}$ and $\ket{111000 ;000111}$. For larger $L$ one will have more types of initial states. It can be shown that, for the half filled sector, the number of available types of initial states is $m+1$ for $L=4m$ or $L=4m+2$. However, when one picks any initial state randomly, different types of states have different probability of being picked. In the thermodynamic limit one would expect the probability of $q=1$ states to be around $50\%$ with the rest of the types having a progressively decreasing probability.
 
The four groups of states seen in Fig. \ref{fig7}(a) are the groups labeled by $q=1,2,3,4$ from top to bottom. States labeled by $q=1$ generally show maximum growth in $\SN$, followed by $q=2,3$ etc. Clearly, if we choose a value of $\SN$ in the plot and study when  each group of states reaches that value, different groups do so in different times, if they reach at all. The reason for this is based on the number of intermediate states in the available Hilbert space an initial state needs to cross to reach the first state with a different particle number in subsystem $A$ ($n_A$). The larger this number, less is the probability and longer the time taken by an initial state to reach a significant value of $\SN$. With increasing $q$ this number increases monotonically. Since we are discussing the MBL phase, where particle movement is suppressed exponentially, this feature is quite prominent. Additionally after the initial rise we also see the groups of states $q=2$ and $q=3$ splitting into two subgroups. This is due to the presence of interactions in the model. The energy cost is different for a particle hopping to a site depending on whether it has an unoccupied or occupied site next to it.

 In Fig. \ref{fig7}(b) we take the numerical data for different configurations for $L=12$ and $W=10$ and find their average after grouping them according to the initial state type. The different colours correspond to different $q$ values. The mean values of $\SN$ at all times show a clear difference with type of initial state at this disorder strength. This points to a strong suppression of any effective long range processes, as higher $q$ values show lower mean $\SN$. In fact the mean values show an almost exponential suppression in the value of long-time $\SN$ with increasing $q$, which is what one would expect in the MBL phase.
 
 Seeing how $q=1$ states make up for the bulk of mean $\SN$ we will try to analytically predict short time behaviour for these states using a 2-state model. Note that states with higher $q$ values need at least a $q+1$ state model to explain their behaviour in short times.
\subsection{Analysis via two-state model}
\begin{figure}
\includegraphics[width=0.8 \columnwidth]{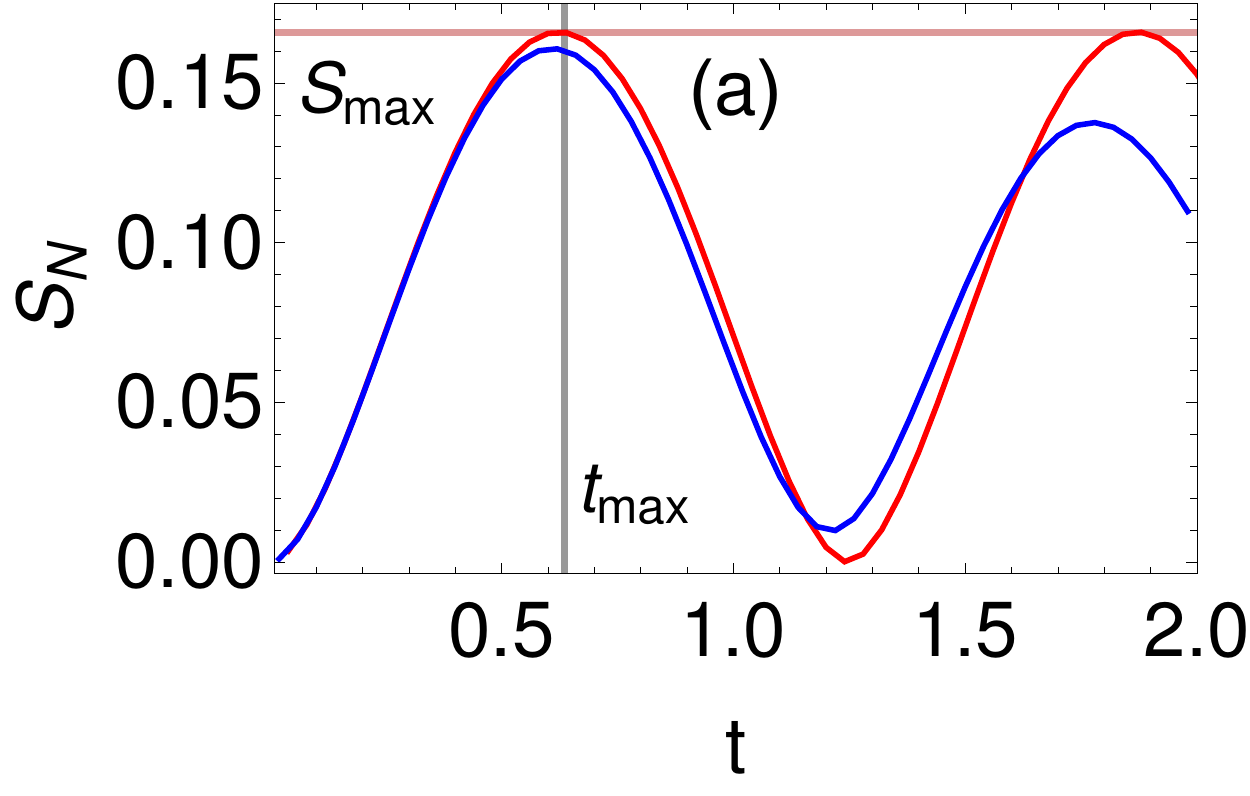}
\includegraphics[width=0.8 \columnwidth]{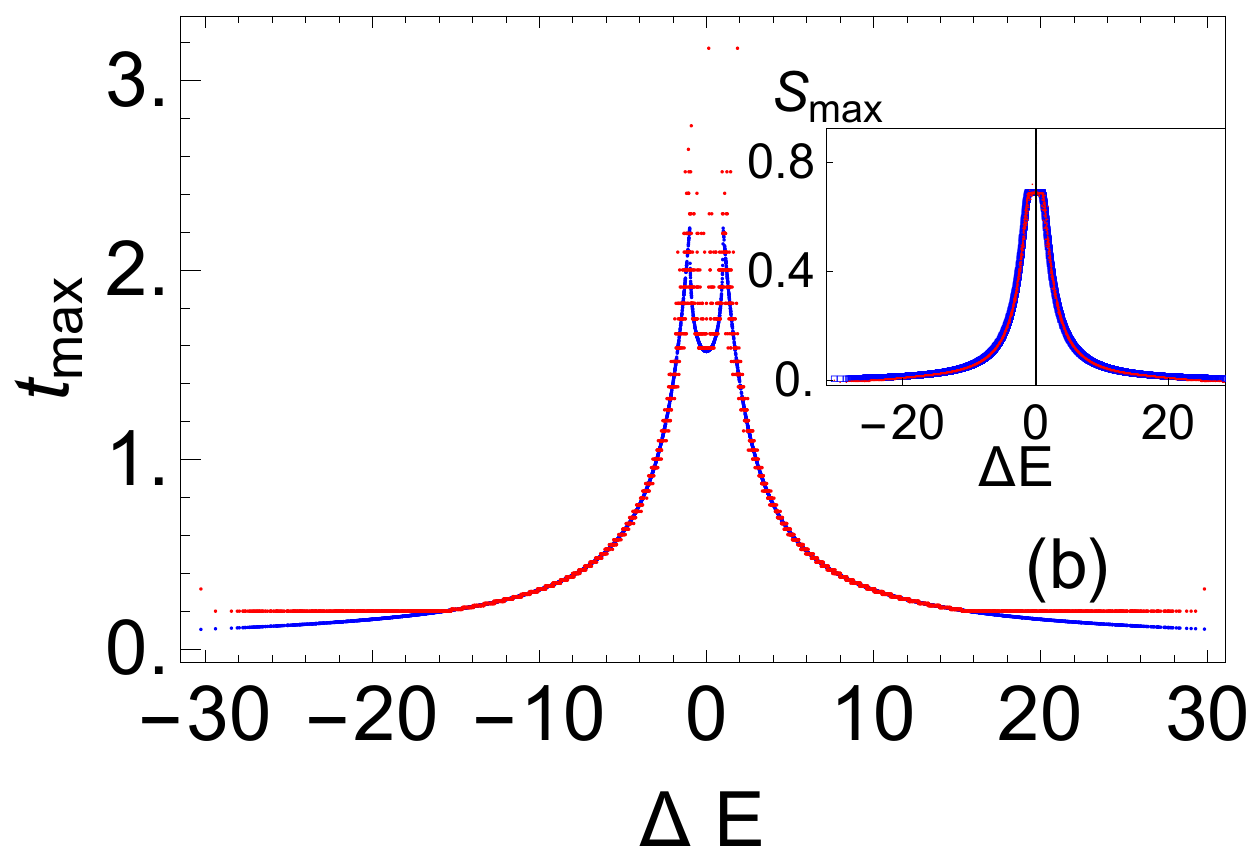}
\caption{Plots showing agreement of results from two-state model [Eq. \ref{twostate}] with exact numerics  for $W=15$ and $L=12$. (a) Comparison of analytical results (red line) from Eq. \ref{SNexp} with exact numerics (blue line) for one single configuration. (b) Plot showing the agreement of numerical (blue dots) with analytical (red dots) data obtained from Eq. \ref{part1} for $t_{max}$.  The inset shows agreement of data for $S_{max}$.}
\label{fig7c}
\end{figure}
The rise of $\SN$ in the $q=1$ set of states can be approximated via a two-state model from initial state $\ket{1}$ (with diagonal term $E_1^0$ in the Hamiltonian) to a state with one hop across the boundary $\ket{2}$ (with diagonal term $E_2^0$) with a Hamiltonian again given by Eq. \ref{twostate}.
Recall the amplitude and frequency of oscillation in the two-state problem is given by Eq. \ref{amp} and \ref{freq} discussed in the previous section with $j=J/2$ in this case. To be noted here $E_1^0$ and $E_2^0$ will differ by position of one particle so it generally takes the form $|E_1^0-E_2^0|=|h_{L/2}-h_{L/2+1}\pm J|$ or just $|h_{L/2}-h_{L/2+1}|$ depending on the states $\ket{1}$ and $\ket{2}$. One can extract the position of the first maximum of each of the realizations of $\SN(t)$ from Eqs. \ref{dt} and \ref{SNexp}, which turns out to be at 
\begin{equation}
t_{max}=\begin{cases}
&\cos^{-1}(\frac{A-1}{A})/\Omega , \hspace{0.2in} A>\frac{1}{2}  \\
&\pi/\Omega,  \hspace{0.6in} \text{otherwise}. 
\end{cases}
\label{part1}
\end{equation}
Hence the value of first maxima is given by $S_{max}=\SN(t_{max})$. In Fig. \ref{fig7c}(a) we focus on a single random configuration and show how well our $2$-state model approximates $\SN$ at initial times. We clearly see that $t_{max}$ and $S_{max}$ from the numerical data and that calculated from the two-state model agrees very closely. It is to be noted that the timescales for which our approximation gives very good results is different for different configurations and usually is inversely proportional to $\Delta E =(E_1^0-E_2^0)$. In Fig. \ref{fig7c}(b) we show how well the analytically calculated values of $t_{max}$ and $S_{max}$ matches with the numerical data for $L=12$ and $W=15$.

If $\Delta E$ is small it causes a nearest neighbour resonance, and $A$, the amplitude of oscillation, becomes large which in turn cause $S_{max}$ to have a value close to $\ln 2$. This feature is clearly visible in the inset of Fig. \ref{fig7c}(b). There is a steady growth in $S_{max}$ with decreasing $\Delta E$ peaking at $\ln 2 \sim 0.693$.  In the same plot we show a comparison of the times ($t_{max}$) at  which the first peak occurs, obtained from exact numerics (red points) and the two state model (blue points). In both the plots we find a very good agreement of the analytical prediction with actual numerical data. For large $\Delta E$, $\Omega$ is large and $A<\frac{1}{2}$, hence $t_{max}=\pi/\Omega$, but when $\Delta E$ becomes smaller and makes $A$ become greater than $\frac{1}{2}$ the function changes. From Eq. \ref{amp} and \ref{part1} one can see that the change occurs when $|\Delta E| = \sqrt{3}J \sim 1.72$ with our parameters. This is approximately where we see a sharp change in the nature of the blue curve in Fig. \ref{fig7c}(b). The small deviation between analytical and numerical results at the tails for $t_{max}$ is because for the numerical data the smallest time step taken was $dt=0.2$.

\subsection{N\'eel and Domain Wall state}
\begin{figure}
\includegraphics[width=0.7\columnwidth]{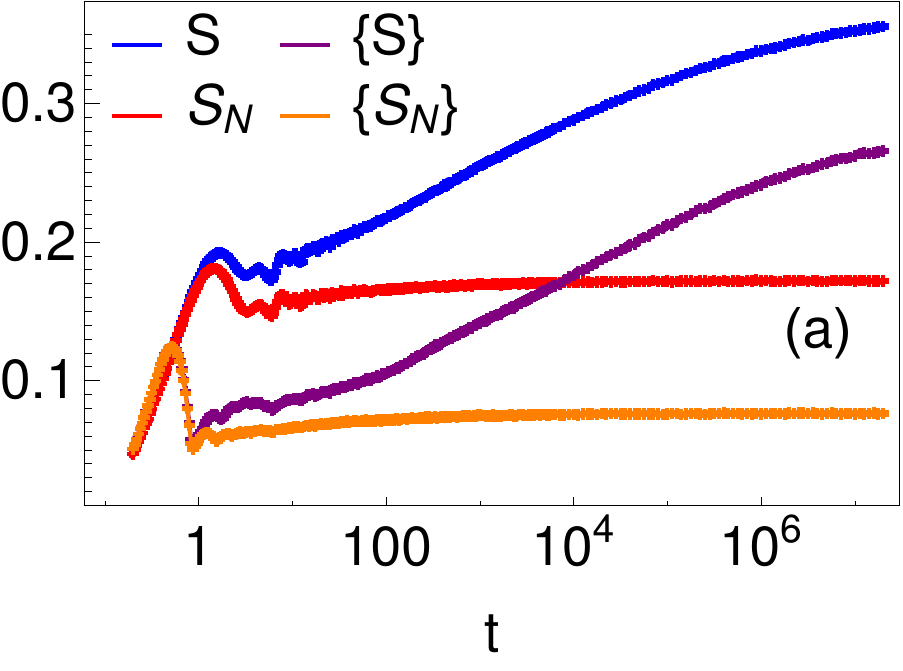}
\includegraphics[width=0.7\columnwidth]{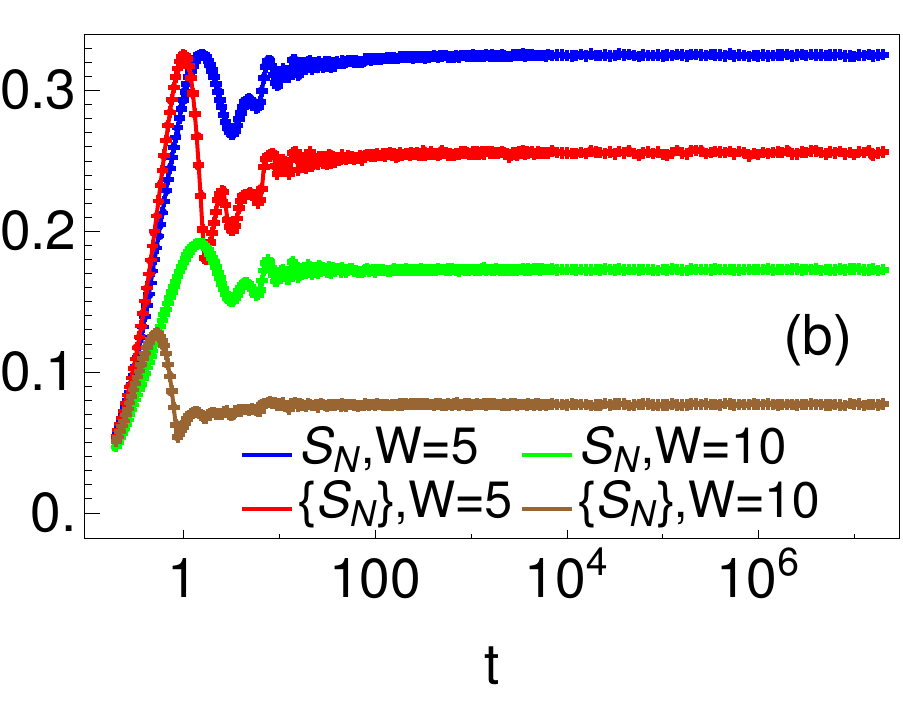}
\caption{Plot showing mean and median growth of entropy for two special states for $L=12$. (a) Comparison of growth of mean ($S$ and $\SN$) and median ($\{S\}$ and $\{\SN\}$) of von Neumann entropy and number entropy respectively vs $t$ at $W=10$ for N\'eel initial state. (b) Growth of $\SN(t)$ for domain wall state.}
\label{fig7a}
\end{figure}

Next, we are going to discuss two very special initial states  which are important in experiments,  whose behaviours are known to be very different from each other and are worthy to be shown separately.\cite{PhysRevB.94.161109,PhysRevB.103.024202} Both fall under the group of initial states $q=1$. One of them is the N\'eel state, which for example is $\ket{10101010}$ for $L=8$.  One expects the $\ln t$ behaviour of von Neumann entropy $S$ to be prominent when the initial state is this state.\cite{PhysRevB.100.125139} A comparison between the behaviour of mean von Neumann entropy $S$ and mean number entropy $\SN$ for this state presented in Fig. \ref{fig7a}(a) shows that while mean $S$ clearly grows logarithmically with time, $\SN$ saturates after the initial growth. In fact $\SN$ never exceeds the first peak at around $t \sim 1$ throughout the entire period. The medians, $\{S\}$ and $\{\SN\}$ also show the same behaviour. 

This result is exactly what one would expect in the MBL phase, where after an initial growth of $\SN$ which follows the growth of $S$, $\SN$ will saturate. Since in the MBL phase one expects transport of particles to be heavily suppressed, one expects no growth of $\SN$ after initial rise within localization length. However configuration entropy, which forms the other part of von Neumann entropy can still grow due to dephasing, which does not require actual particle transport across the boundary. This growth requires interactions which can entangle pairs of particles and thus entangles particle configurations across the boundary. Hence one expects $\SN$ to saturate but $S_{\rm{conf}}$ to grow indefinitely in the thermodynamic limit which in turn causes $S$ to grow. The small amount of growth in $\SN$ seen in the plot at times $t<10^3$ is due to the resonant disorder cases which will be further elaborated in the next section.  Furthermore, the fact that $\SN$ does not cross the maximum of the first oscillation, which has been shown to be described by one particle crossing the subsystem boundary to the neighbouring site, indicates that the particles do not travel much farther in the lattice than that, and the rest of the entropy $S$ is given by configurational correlations between the subsystem and the environment.

The other state which is important is the domain wall state, which for example is $\ket{11110000}$  for an $L=8$ system. By choosing this state we drastically reduce the local connectivity of the relevant region of Hilbert space, and thus one would expect very few resonant cases to be present if one starts from this state. If the particle transport at long distances is indeed being exponentially suppressed by localization, the resonances will only occur near the boundary of the subsystem. Lack of long-range resonances mean the growth of $\SN$ should occur only during short times before saturation. In  Fig. \ref{fig7a}(b) we see exactly this expected result. The value of $\SN$ saturates pretty quickly($t < 100$) for $W=5$ and increasingly faster with increasing $W$. As in the N\'eel state, in this range of disorders, $\SN$ never goes beyond the first peak meaning there are almost no cases where resonance with more than two states are involved. Reduction in Hilbert space connectivity for the domain wall state also causes $\SN$ to reach a saturation value much earlier for this state than the N\'eel state which is also visible in Fig. \ref{fig7a}.

\begin{figure*}
\includegraphics[width=0.7 \columnwidth]{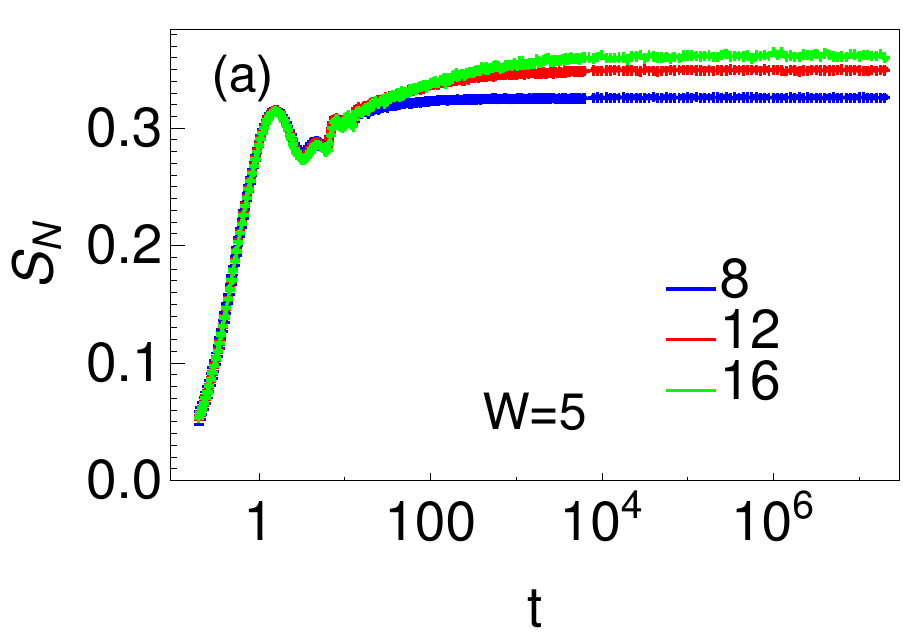}
\includegraphics[width=0.7 \columnwidth]{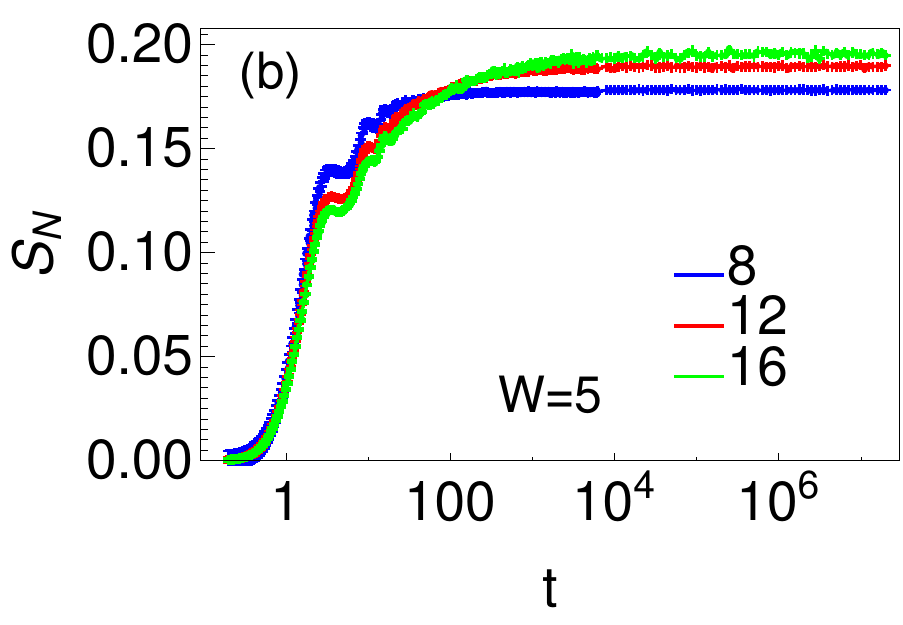}\\
\includegraphics[width=0.7 \columnwidth]{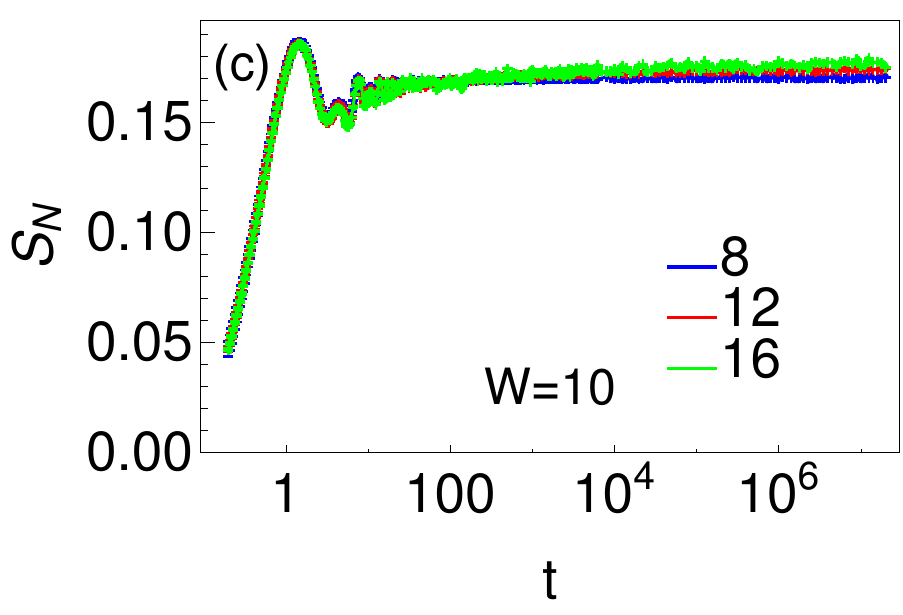}
\includegraphics[width=0.7 \columnwidth]{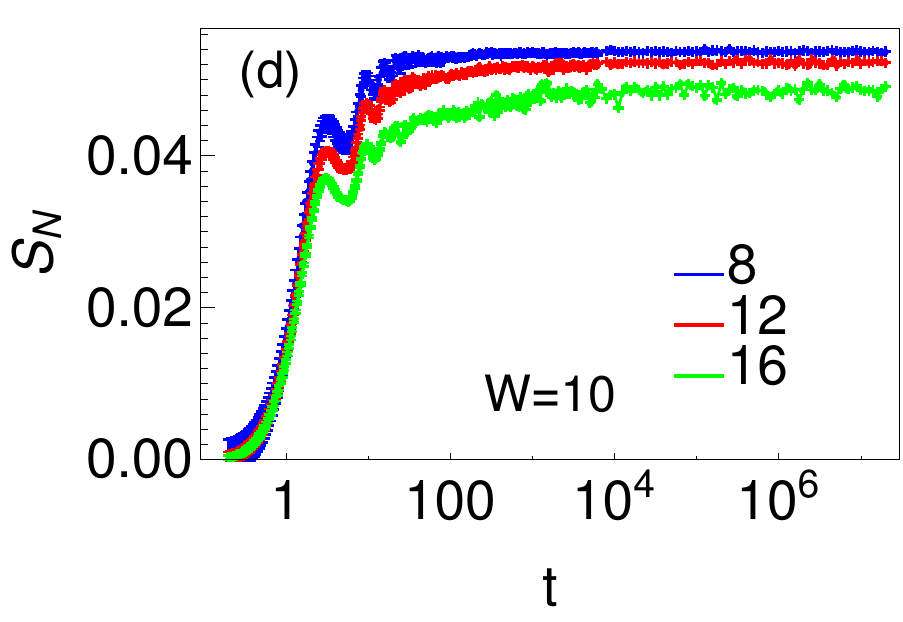}
\caption{Plot showing how $q=1$ states [(a) and (c)] and $q \neq 1$ states [(b) and (d)] show different growth of mean $\SN(t)$ for different $L$ (indicated by different colours in the plot).}
\label{fig8a}
\end{figure*}
\subsection{Contrasting behaviour of different initial states}
At high disorder strengths we have seen in Figs. \ref{fig1} and \ref{fig9} that $\SN$ shows a tendency of decrease with increasing system size. In Fig. \ref{fig1} this trend already shows up at $W=10$ where the mean $\SN$ for $L=16$ is lower than for $L=8$. We have also discussed in Sec. \ref{steady} that the behaviour of steady state $\SN$ with system size show a reversal of trend beyond a certain disorder strength, when it actually decreases with increasing $L$. This puzzling behaviour also reported in Ref.~\onlinecite{PhysRevB.103.024203} has its source in the way we choose our initial states. Since different initial states show markedly different growth of $\SN$, averaging over all of them results in this behaviour.

To elaborate our point, in Fig.\ref{fig8a}, we separate out $q=1$ and other types of initial states for disorders of strength $W=5$ and $W=10$, two strengths which show opposing trends. One clearly sees if one uses only $q=1$ initial states, as in Fig. \ref{fig8a} (a) and (c), $\SN$ shows an increase with system size for both $W=5$ and $W=10$, with significant suppression of growth at $W=10$. It is also important to note how the increase of long-time $\SN$ for $W=5$ when one goes from $L=8$ to $L=12$ is significantly greater than when one goes from $L=12$ to $L=16$, which hints at non-ergodicity. At $W=10$ this growth is even smaller.  In Fig. \ref{fig8a}(b) and (d) we then show that when we take the mean over all the other kinds of states (which still make up for approximately $50\%$ of the initial states), one sees a clear decrease at $W=10$ with increasing system size, while for $W=5$ there is small increase.

To explain this observation, recall from Fig. \ref{fig7}(b) and Sec. \ref{resonant} that larger disorder strengths around $W \ge 10$ inhibit long range resonances and long range particle transport. Hence mean $\SN$ at long times shows a progressive decrease in magnitude with increasing $q$. As we increase $L$, the types of computational states available in the pool of initial state increases. The presence of states with higher $q$ value in the pool in turn means increase in number of configurations where $\SN$ has a lower value. Hence when one takes the mean over all such configurations, redistribution of weights towards lower values of $\SN$ for $q \neq 1$ states with increasing $L$ will cause a decrease in mean $\SN$. Since at $W\sim 5$, some long range processes are still allowed, one can still see significant growth of $\SN$ for initial states with larger $q$ values. This in turn shows up as an increase in $\SN$ with system size at $W=5$ for both $q=1$ and $q\neq 1$ states as seen in Fig. \ref{fig8a} (a),(b). But as expected this feature disappears for larger disorder as seen in the $W=10$ case presented in Fig. \ref{fig8a} (c),(d).

This analysis also explain the reversals of trend of $\{\bar{S}_N\}$ with system size as we increase disorder shown in Fig. \ref{fig9}(b). Beyond a disorder strength, due to suppression of long range resonance, configurations with low value of steady state $\SN$ increase with increasing $L$. Taking mean and median over such distributions result in the behaviours seen.

 \section{Dynamics at intermediate times}
 \label{inter}
In this section we shall discuss the dynamics of the system at intermediate times. First, we shall show that resonant configurations, discussed previously in Sec. \ref{resonant}, account for most of the growth seen in $\SN$. Then we shall show how we can fit the growth of $\SN$ in a power law fit and provide an explanation for it from the resonance picture.
\subsection{Filtering out resonant configurations}
In Sec. \ref{model} we discussed how averaging over time evolution of $\SN$ containing `resonant' configurations, with resonance occurring at varying $t$ shows up as a slow growth in mean or median $\SN$.    To show that the resonant cases are indeed responsible for the slow growth in $\SN$, ideally one should be able to filter out these cases and obtain a perfectly non growing curve. In practice however detecting all the resonances is notoriously difficult because they have a wide variety of strengths which all contribute to the small growth in $\SN$. Hence, we will filter out the considerably large jump cases, where we consider any sudden rise of height of $\SN$ denoted by $\delta S$ $> 0.1$ as a `resonance'. For a pictorial description of identification of resonances refer to Appendix \ref{appC}. More exotic filtering procedures can be developed but we have verified that they do not make any qualitative change to the results presented here.

\begin{figure*}
\includegraphics[width= 1.5 \columnwidth]{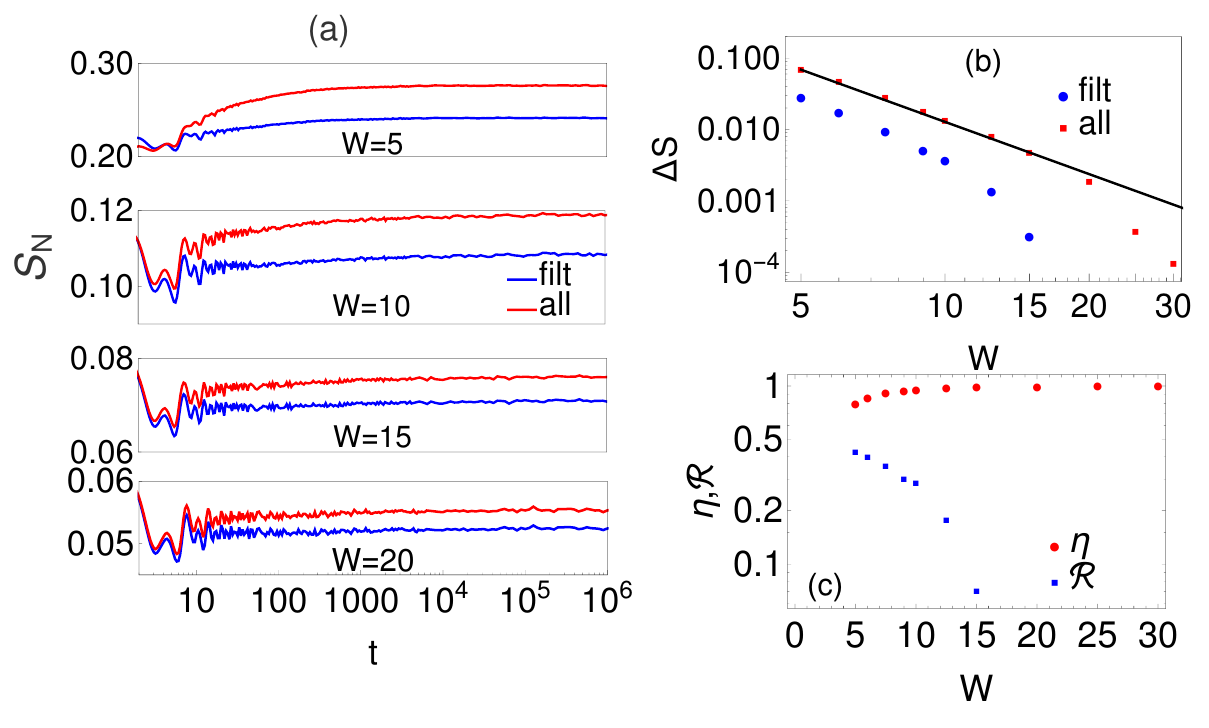}
\caption{Comparison of resonance filtered vs unfiltered data for $\SN(t)$ at $L=12$. (a)Plots of data before (red) and after (blue) resonance filtering for different disorder strengths $W$.
(b)Plot showing a comparison of the increase of $\SN$ denoted by $\Delta S$ from initial times to late times for all data (red), and the data where resonant configurations are filtered out (blue). See text and Appendix \ref{appC} for details. Black line denotes fit $W^{-1.62}$. (c) The red dots show the fraction of realizations $\eta$ which do not have a jump, $\delta S$ higher than the tolerance of $0.1$, while the blue dots denote the ratio $\mathcal{R}$  between the two sets of data on the above panel. }
\label{fig4}
\end{figure*}
With this tolerance, Fig. \ref{fig4}(a) shows the difference between mean $\SN$ in situations where we filter out the jumps vs the full data. We do see a marked decrease in the difference between $\SN$ at initial times and late times for the two sets of data denoted by blue and red.  However since the tolerance chosen is arbitrary one cannot expect plots to become completely flattened out after filtering. The filtering improves as we increase disorder. This is expected as at lower disorder where localization length is larger, the eigenvectors have longer tails which adds to particle transport beyond one-site but with exponential suppression with distance, and there are weak resonances connecting large distances in the system. These result in jumps which are smaller than the tolerance value chosen but many in number, which are difficult to filter out due to the oscillatory nature of $\SN$.

In Fig. \ref{fig4}(b) we try to quantify the amount of growth our filtering process takes away from $\SN$. To do this we consider the difference ($\Delta S$) between $\SN$ averaged over late times $10^6 -10^7$ ($\tilde{S}_N$) and at initial times. Since $\SN$ is highly oscillatory at initial times, we take the average value of $\SN$ between the first two peaks of the data, i.e. in the orange region in Fig. \ref{fig3b} in Appendix \ref{appC}. We plot this quantity for the two sets of data, comparing the set with all configurations with the set where we remove the resonant configurations. Clearly we see with increasing disorder our filtering process becomes more accurate in removing the $\Delta S$ seen, which itself decays as a power law with $W$ for intermediate disorder, and faster for larger disorders. The fact that $\Delta S$ is such a small quantity and can almost be completely removed when we filter out the resonances clearly points to lack of ergodicity in the system. In Fig \ref{fig4}(c) we also see how the number of cases to be filtered decreases quickly with increasing $W$ thus clearly showing how a very small fraction of cases contribute to the slow growth in entropy seen. It shows that while the ratio of unfiltered to total cases ($\eta$) quickly tends toward $1$, the ratio ($\mathcal{R}$) between increase in $\SN$ for unfiltered and total cases quickly tends toward $0$, confirming a small number of resonant cases are responsible for the increase (Also see Appendix \ref{anderson} for a comparison with Anderson localized systems).

\begin{figure}
\includegraphics[width=0.8 \columnwidth]{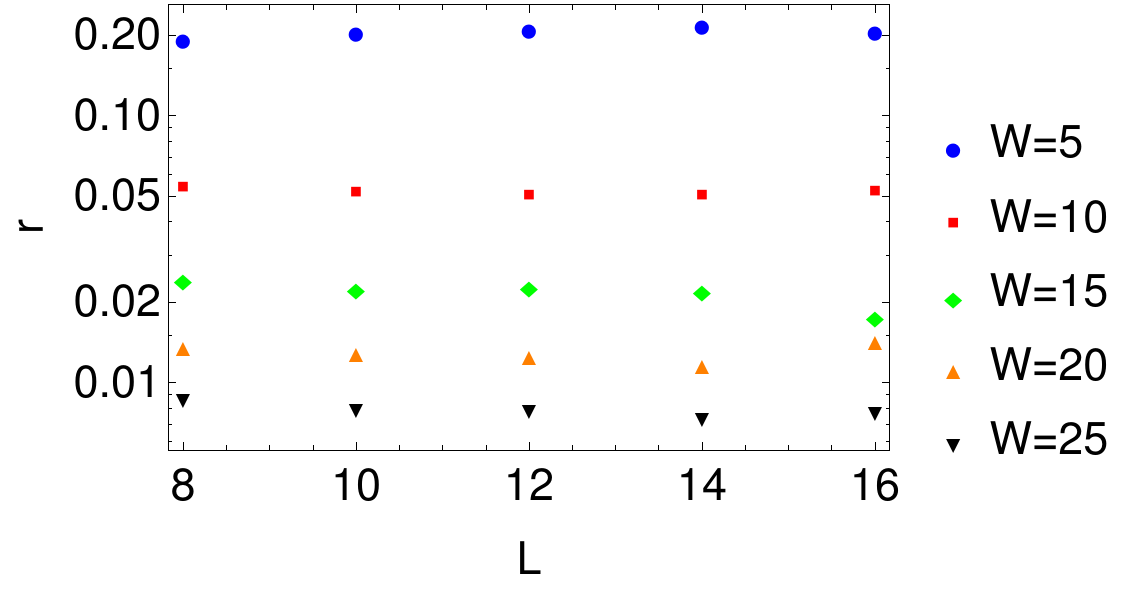}
\caption{Plot showing fraction of total number of cases ($r$) that can be recognized as ``jumps" for different system sizes and different disorders.}
\label{fig6}
\end{figure}
Finally in Fig. \ref{fig6} we see how many states get filtered out for different disorder strengths at different system sizes using the same filtering procedure. $r$ denotes the ratio between number of resonant cases to the number of total configurations, considered for different $L$ and $W$.  We observe that while with increasing disorder the cases which need to be filtered out fall off sharply, there is almost no change with system size. The fact that there are no signatures of growth of fraction of resonant cases with system size, shows further that the system is not ergodic. However, do note that we do not make any distinction between short range and long range resonances for this result. It can be interesting to do a closer comparative study of range of resonances in the framework of $\SN$ growth times, in a manner similar to Ref.~\onlinecite{david21} and this is left for a future work.

\begin{figure}
\includegraphics[width=0.7\columnwidth]{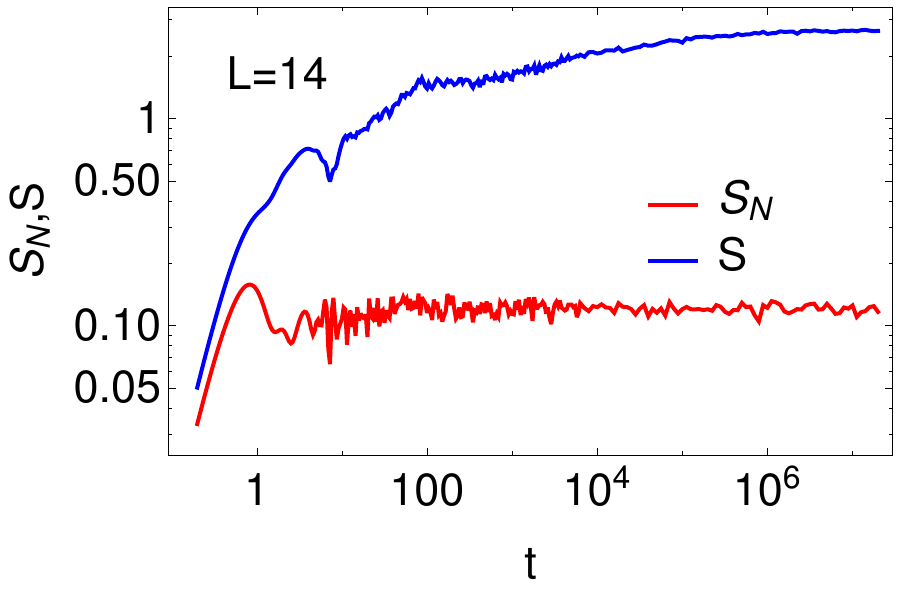}
\caption{Plot showing comparison of growth between $S(t)$ and $\SN(t)$ starting from state $\ket{\mathcal{I}}$ for one typical disorder realization with $W=10$.}
\label{fig11a}
\end{figure}
\subsection{Power law growth of $\SN$}
\begin{figure*}
\includegraphics[width=0.6\columnwidth]{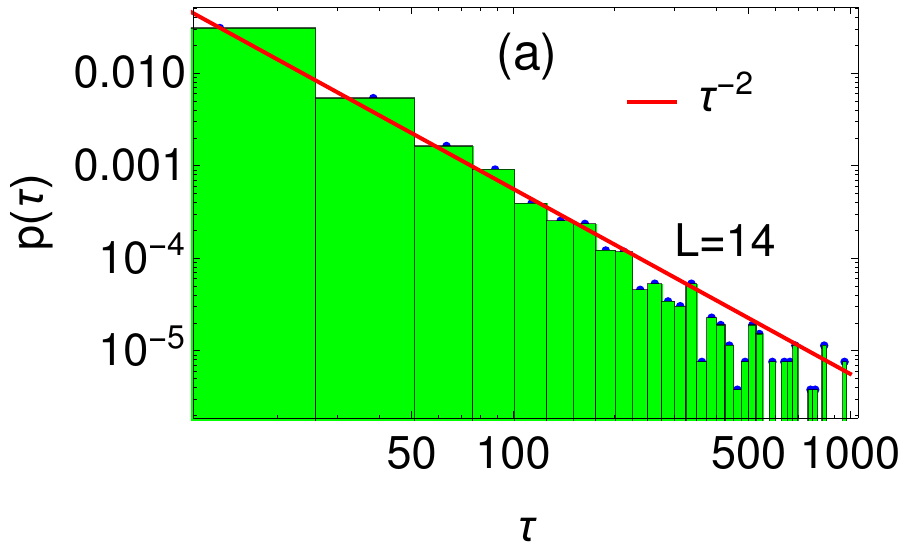}
\includegraphics[width=0.6\columnwidth]{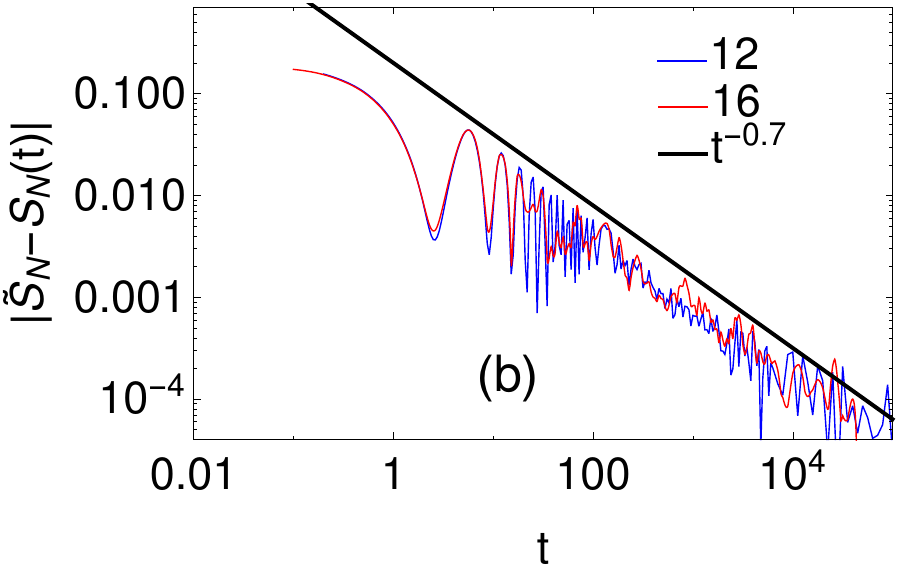}
\includegraphics[width=0.6\columnwidth]{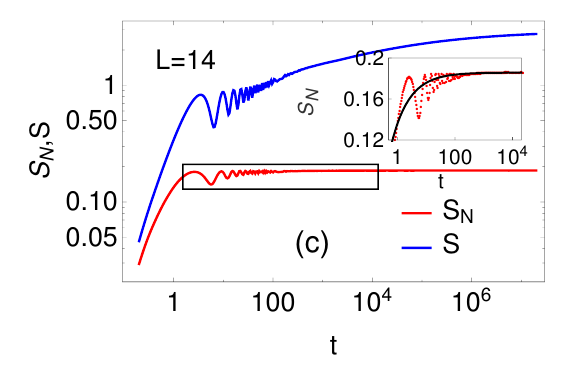}
\caption{Analysis of behaviour of $\SN(t)$ for initial state $\ket{\mathcal{I}}$ for $W=10$. (a) Probability distribution of jump times $\tau$ showing how the power law is very close to $\tau^{-2}$.  (b) Plot showing approach of mean $\SN$ toward its long-time average value $\tilde{S}_N$ for two system sizes denoted by blue and red.  (c) A comparison of how mean von Neumann entropy $S$ grows with time vs disorder-averaged $\SN$. Inset: Plot zooming in on the area marked by black rectangle in the main plot. The black line denotes the power law fit of the form $A-B/t^{0.7}$.}
\label{fig11}
\end{figure*}

We will end this section by showing the law of growth of $\SN$ during intermediate times numerically, since $\SN$ in this timescale is very difficult to treat analytically. However, we can have an approximate idea of the growth from the distribution of times of the `jump'. Denoting the approximate time of the jump by $\tau$, from similar arguments discussed in Sec. \ref{steady}, one can find the probability distribution of $\tau$ from inverse of the difference between two random numbers, $h_1$ and $h_2$, picked from a uniform distribution. Denoting $z=h_1-h_2$ and approximating $\tau=1/z$, one can show via a simple variable transformation that $p(\tau=1/z)\sim B/\tau^2$. Under the assumption that all the jumps are of equal strength $\delta S$, one can approximate the growth of $\SN$ with time as
\begin{equation}
\SN(t)=\delta S\int_{t_0}^t B/\tau^2 d \tau=\delta S(B/t_0-B/t)
\label{SNcum}
\end{equation} 
where $t_0$ is the approximate lower cutoff timescale. It is chosen approximately as the time after which the resonances which occur are beyond nearest neighbour. Based on our analysis in the last section we set it to be $\sim 1$. We remark that a power-law behavior coming from resonances is observed also in the decay of correlation functions~\cite{anushya20}.

We verify our theory in Figs.\ref{fig11a} and \ref{fig11}. To do so we use the initial state $\ket{\mathcal I}=\ket{\alpha}\otimes\ket{\beta}$ where $\ket{\alpha}$ is an equal superposition of all computational states of subsystem $A$ with $\langle\alpha|N_A|\alpha\rangle=[L/4]$ and $\ket{\beta}$ is an equal superposition of all computational states of $B$ with $\langle\beta|N_B|\beta\rangle=L/2-[L/4]$. $[x]$ denotes greatest integer less than or equal to $x$. This is a more unbiased choice than random computational state and introduces another layer of averaging which smoothens the plots. This allows us to more accurately look at the law governing growth of $\SN$.

In Fig. \ref{fig11a} we first show a comparison of $S$ and $\SN$ for a typical disorder realization of strength $W=10$ for this initial state. We clearly see that while von Neumann entropy $S$ shows a growth for a long timespan, $\SN$ shows saturation at extremely early times. This shows configurational changes to entropy is what drives the $\ln t$ growth of $S$ at late times. In Fig. \ref{fig11}(a) we verify that the probability distribution of jumps with $\delta S$ greater than $0.1$ for $L=14$ and $W=10$ indeed show the power law of $\sim 1/\tau^2$ as expected from our analysis of resonant configurations\cite{Note5}. In  Fig. \ref{fig11}(b) we plot the approach of $\SN$ toward its long-time-averaged value $\tilde{S}_N$ for two different system sizes. We see the approach follows a power-law behaviour given by $1/t^{0.7}$ for around $3-4$ orders of magnitude of $t$ before statistical fluctuations becomes prominent. Note that the agreement is for a longer range of $t$ than if one fits via the $\ln \ln t$ function (also for $W=5$; data not shown) reported in Ref.~ \onlinecite{PhysRevB.103.024203}. We have averaged over $10^5$ configurations for $L=12$ and $2 \times 10^4$ configurations for $L=16$ to obtain this plot. In Fig. \ref{fig11}(c) we confirm that disorder averaged $\SN$ saturates much earlier than mean $S$ for $\ket{\mathcal{I}}$. This shows that in general $S(t)$ and $\SN(t)$ are not related (as opposed to non-interacting systems\cite{sirker02}). In the inset of the same plot we show the comparison of power law fit to mean $\SN$, using the exponent obtained from Fig. \ref{fig11}(b) and we see that it agrees very well. 

The fact that the power law exponent by which $\SN$ approaches $\tilde{S}_N$ obtained from the fit is not exactly $1$ is because of several factors. The assumption of equal jump heights is oversimplified as strength of hybridization is not the same for all cases. Then, the integral approximates the jump via step functions, while in reality these should be oscillatory, and $\SN$ for individual configurations is also a highly oscillatory function. Additionally, different categories of initial states show different behaviour of growth of $\SN$, and while we have managed to reduce this effect by taking a superposition of many states, some signatures still remain. It is to be noted that this choice of initial state allows the early saturation of $\SN$ to be more clearly visible than for a random computational state. This is because to obtain a significant resonance for state $\ket{\mathcal{I}}$ a large fraction of the superposing computational states have to be resonant. For long-range resonances causing growth of $\SN$ at later times this has a much lower probability than a single computational state being long-range resonant. This makes the choice the more generic one for study of growth of number entropy.

\section{Discussion}
\label{disc}

In this work we have discussed in details the source of the slow growth in number entropy $\SN$ in the MBL phase of the disordered isotropic spin-$1/2$ Heisenberg chain. We have shown that there are special initial configurations of disorder realization and initial states, which causes a strong hybridization between two (or more) computational states which differ in number of particles in subsystem $A$. This shows up as a sudden jump in $\SN$ at characteristic times given by the inverse of the energy difference between these two states. When the mean over different configurations is taken, such jumps at different characteristic times are averaged over and show up as a small growth in $\SN$. Looking at the distribution of jump times, the numerics agrees with theoretical prediction, which also suggests a power-law growth of $\SN$. A power-law indeed describes the growth of $\SN$ rather well, though with a power that slightly differs from the one obtained from the simple two-state resonance model.

Furthermore in the system sizes available to us we have discussed the steady state behaviour of $\SN$ showing how mean and median of the quantity decrease with $W$ approximately as $W^{-1}$ and $W^{-2}$, respectively. We have then provided an analytical explanation of this behaviour from the two-state model, further solidifying our two-state hybridization hypothesis. We have also shown how at lower disorder strengths one can indeed see a slight increase in $\bar{S}_N$ with system size $L$ but the trend is reversed for higher disorders, which rules out long-range transport at larger disorder. We then discussed how the initial computational states can be grouped on the basis of their initial time growth and how this affects growth of $\SN$ at all timescales. Finally, we show how by filtering out the large resonances one indeed can get a non-increasing $\SN$ at intermediate times.  

Our results in this work agree with the current picture of MBL. While we do not fully understand the change in behaviour of steady state $\SN$ with $L$ when we go from $W \lesssim 6$ to higher values, we note that it is consistent with arguments for a rather rich finite-$L$ physics due to resonances~\cite{anushya20,david21}. However to establish an exact connection one needs to do further study. As we have shown, one can figure out what initial state and what disorder configuration one needs to choose to get a resonance from exact numerics; replicating a similar setup would provide a good test-bed for verification. Alternatively, if one excludes the configurations where one sees such sudden jumps, one will be able to get a non-growing $\SN$ curve for MBL systems in experiments, which would also serve as a validation to our theory.

Natural extensions of this work would include studying the situation in quasiperiodic potentials and in a Floquet-MBL setup\cite{abanin,PhysRevB.105.024301}. In quasiperiodic systems since the disorder is deterministic, it should be easier to predict the resonances by studying the structure of potential itself. A careful study then might shed light on the differences in the phenomenon of localization in interacting quasiperiodic systems and systems with random disorder. Since Floquet Hamiltonians are long-ranged near the Floquet-MBL critical point (which is the frequency of the external periodic drive at which the ergodic to Floquet-MBL transition occurs) compared to short-ranged Hamiltonians of ordinary MBL, one can naively expect such resonant cases to be more prominent there and it warrants a careful study. 

As a final note, we remark that these results show that the unitary transformation $U$ rotating l-bits to a computational basis (i.e., $U$ that consists of eigenvectors) is the central object that one needs to study in order to understand the localized phase. Localization is not so much about properties of eigenvalues (which do not tell us anything about spatial extent of eigenfunctions) but more about properties of $U$, specifically, $U$ has to be quasi local. This is explicit in proofs of MBL that perturbatively start from the infinite disorder limit\cite{Imbrie16}, where the central point is to have control over possible resonances in $U$, as well as in our theory of the number entropy growth where accounting for resonances is crucial.
Indeed if one tried to approach the problem using the local integrals of motion (LIOMs) model~\cite{PhysRevB.90.174202,PhysRevLett.111.127201,serbyn14,anto15} for MBL by replacing the exact eigenvectors connecting the model-dependent LIOM basis to the computational basis via a generic unitary matrix built up by local quantum gates, one would not be able to replicate the results (See Appendix \ref{LIOM} for details).

 In this sense $\SN$ is also the more precise probe of localization than the von Neumann entropy. The number entropy explicitly focuses on localization -- it needs to be bounded if one has localization -- whereas in the von Neumann entropy a constant contribution due to quasi-local $U$ is masked out by the (more trivial) logarithmic growth due to dephasing.\\

Support from Grants No.~J1-1698 and No.~P1-0402 from the Slovenian Research Agency is acknowledged.


%

\appendix

\section{Effective Hamiltonians for resonances}
\label{effham}
We analyzed how one can approximate growth of $\SN$ in resonant configurations via few-state models in Sec. \ref{resonant} for special initial states. However, as we remarked there, for a generic computational state, the method developed is not very good due to involvement of many intermediate states. In this appendix we shall demonstrate two methods by which we can approximate the growth in $\SN$ via a low-dimensional subspace for any initial resonant state.

\begin{figure}
\includegraphics[width=0.7\columnwidth]{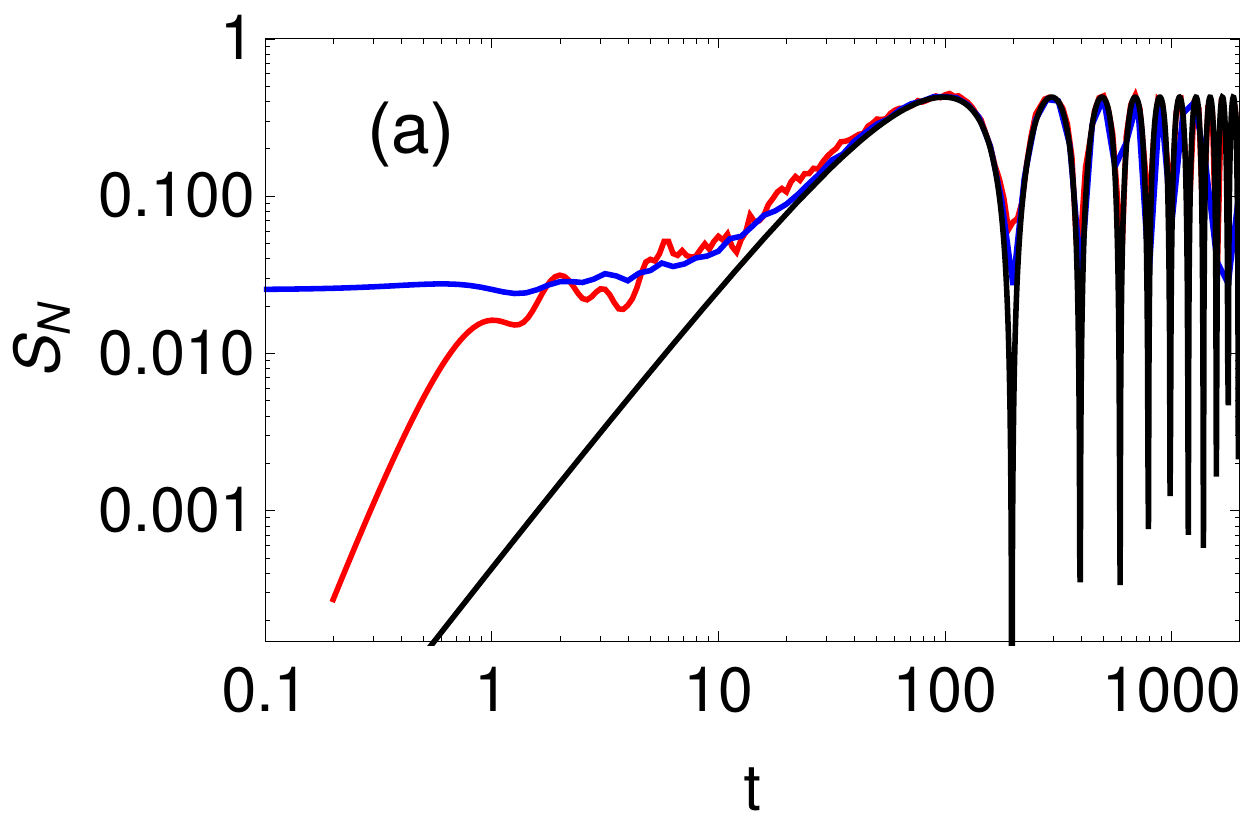}
\includegraphics[width=0.7\columnwidth]{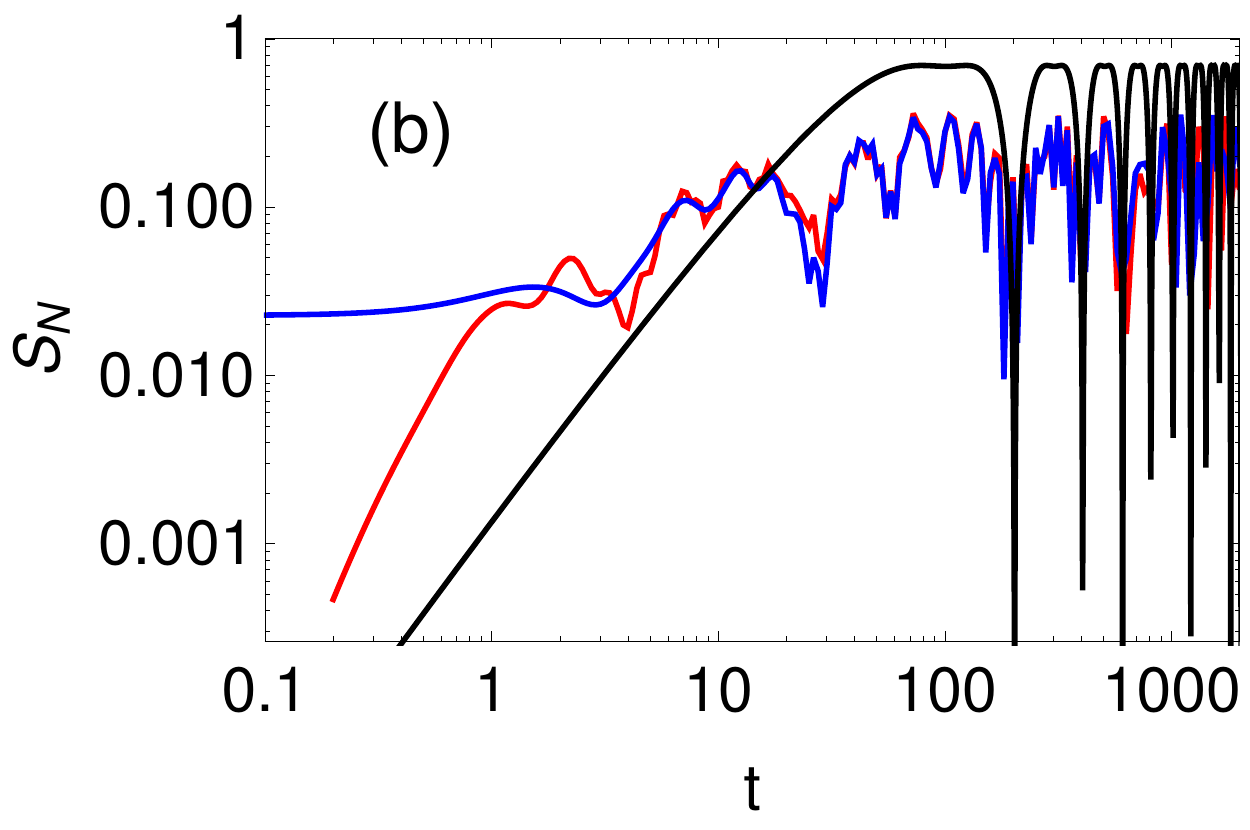}
\caption{ Plots showing how effective low dimensional Hamiltonians capture the later rise in $\SN(t)$ which is caused due to resonant states. The red line denotes result from exact numerics, the black line denotes the result from the effective $2$-body Hamiltonian, and the blue line denotes the result after projecting the initial state into an effective $s$-dimensional subspace at $W=10$ and $L=12$. (a) Plot showing a single resonant case where $s=3$ is enough. (b)Plot showing a case  which supports $2$ resonances. $s=5$ is needed in this case}
\label{figapp1a}
\end{figure}

The first method we discuss is the one in Ref.~\onlinecite{david21}. In this scenario, one needs to diagonalize the Hamiltonian to identify the two relevant resonant eigenstates. To do so one needs to look at two eigenstates having small energy difference but largest overlaps with the initial computational state. Now after writing a $2 \times 2$ matrix in this subspace, with the diagonal elements given by the eigenenergies, one rotates back to a local basis using an optimization technique to create an effective Hamiltonian which can describe the resonance. In Fig. \ref{figapp1a} we take two example cases and show what the effective Hamiltonian gives us. This process is not fully analytical as we still need to identify the relevant resonant eigenfunctions; nevertheless it serves as a demonstration of how the system can be described by an effective low-dimensional model. The result from this method given by the black line in panel (a) is extremely accurate at capturing the late rise of $\SN$ due to the resonance. Since we have only two states involved in this process one would see Rabi oscillations with the time period inversely proportional to the eigenenergy difference.
  
One can also approach this problem in a different manner. Since we are in the MBL phase, one can assume that the initial computational state will show a significant overlap with very few eigenstates. For resonant behaviour we know the strength of overlap has to be significant, so for these cases one can work in the subspace of $s$ eigenvectors where $s\lesssim 5$. Hence we take the $s$ largest overlaps of the initial state with the eigenvectors, and renormalize to get an effective initial state in this subspace of $s$ eigenvectors and proceed with the calculation. The advantage of this over the previous method is, since more than $2$ vectors are involved, the oscillations will show more features and will be closer to the actual $\SN$. Additionally, there are some configurations where there are multiple resonances; i.e., instead of a single jump, $\SN$ shows more than one jump occurring at different timescales. This can happen if the initial state has significantly large overlaps with several eigenvectors with energies very close to each other. These are quite rare and almost measure zero events deep inside the MBL phase. For this rare multiple-resonance scenario, one will still be able to capture the physics of $\SN$ within this small subspace. However one should remember that analysis of the effective Hamiltonian would be purely numerical for $s>4$ and since the effective initial state will not be a product state, the initial $\SN$ will be a small non zero value.
 
These features can clearly be seen in Fig. \ref{figapp1a}. In Fig. \ref{figapp1a}(a) we show the situation for a configuration which has a single resonance. While the effective $2$-state model is able to approximate the jump very well, the $3$- eigenstate subspace correctly captures the more intricate features of $\SN$. In Fig. \ref{figapp1a}(b) we choose to show a double resonant configuration where the $2$-state model cannot accurately capture the scenario, and we require a subspace of $5$ eigenstates to get an almost accurate approximation of $\SN$. Notice how for the eigenstate subspace case $\SN$ starts from a small non-zero value at small $t$ as the initial state now slightly deviates from a product state.  This analysis further strengthens our claim of how these rises occur due to resonances, which can be described by an effective low-dimensional theory, removing any notion of presence of ergodicity.
\section{Distribution of $\bar{p}$}
\label{appB}
In this appendix we elaborate on how we obtain Eq. \ref{powerlaw} from Eq. \ref{psteady} of Sec. \ref{steady} of the main text. In principle, the quantity, $E_1^0-E_2^0$ can have three types of values, $h_1-h_2$ and $h_1-h_2 \pm J$ where $h_1$ and $h_2$ are two random on-site potentials and $J$ is the strength of interaction, depending on which states we are looking at. However we will work in the regime where $J \ll W$, so just considering the case $h_1-h_2$ is enough to obtain the leading behaviour of the quantities we are interested in. Taking random numbers $h_1$ and $h_2$ from a uniform distribution, one can calculate the distribution of $\bar{p}=\frac{2 j^2}{z^2+4 j^2}$, where $z=h_1-h_2$ via variable transformations as
\begin{equation}
w(\bar{p})=\frac{j \left(\sqrt{2} W \sqrt{\frac{\bar{p}}{1-2 \bar{p}}}-j\right)}{2 W^2 \bar{p}^2}, \hspace{0.15 in} \bar{p} \in \left[\frac{2j^2}{4W^2+4j^2},\frac{1}{2}\right].
\label{distrib}
\end{equation} 
One can extract the mean of this distribution as \\
\begin{equation}
\langle\bar{p}\rangle=\frac{j \left(2 W \cos ^{-1}\left(\frac{j}{\sqrt{j^2+W^2}}\right)-j \ln \left(\frac{W^2}{j^2}+1\right)\right)}{2 W^2}.
\end{equation}\\
Taking $W\gg j$ and expanding the series in $1/W$, we get
\begin{equation}
\frac{\pi  j}{2 W}+O(1/W^2),
\end{equation}
where the leading term goes as $1/W$.
For the median one has
\begin{equation}
\{\bar{p}\}=\frac{1}{\frac{\left(3-2 \sqrt{2}\right) W^2}{j^2}+2},
\end{equation}
which again can be expanded in $1/W$ as, 
\begin{equation}
\{\bar{p}\}=\frac{\left(2 \sqrt{2}+3\right) j^2}{W^2}+O(1/W^4)
\end{equation}
which goes as $1/W^2$ for $j \ll W$.
\section{Additional results for the steady state}

\label{appB1}
\begin{figure*}
\includegraphics[width=0.66 \columnwidth]{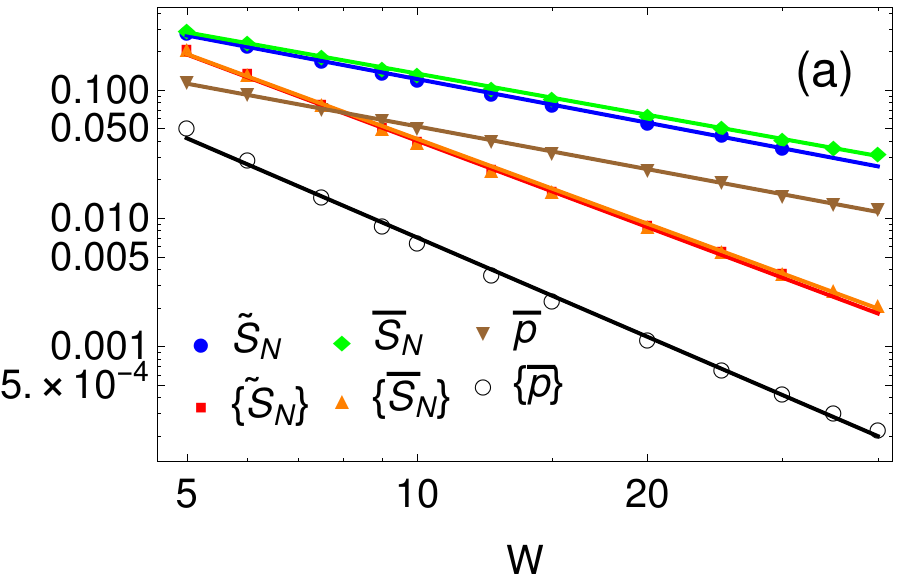}
\includegraphics[width=0.66 \columnwidth]{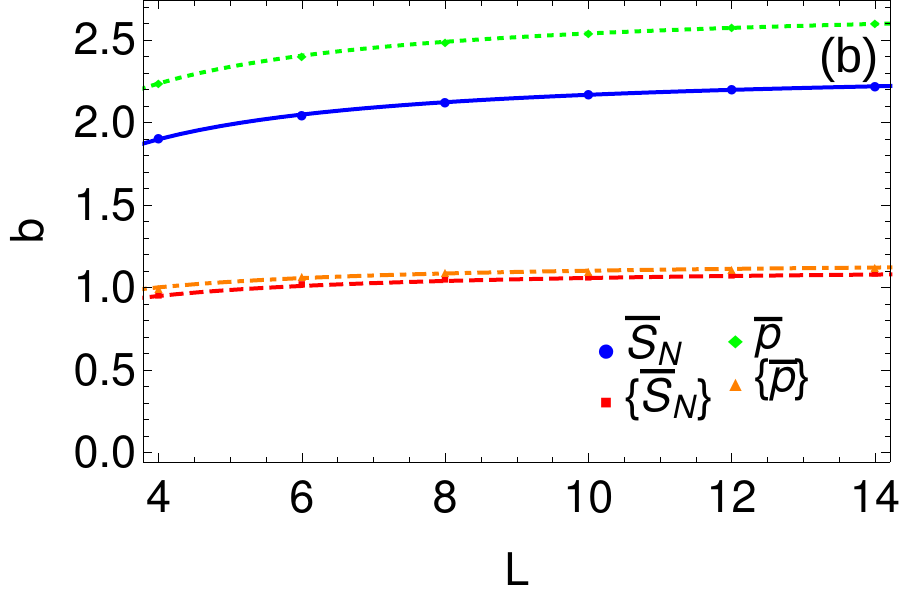}
\includegraphics[width=0.66\columnwidth]{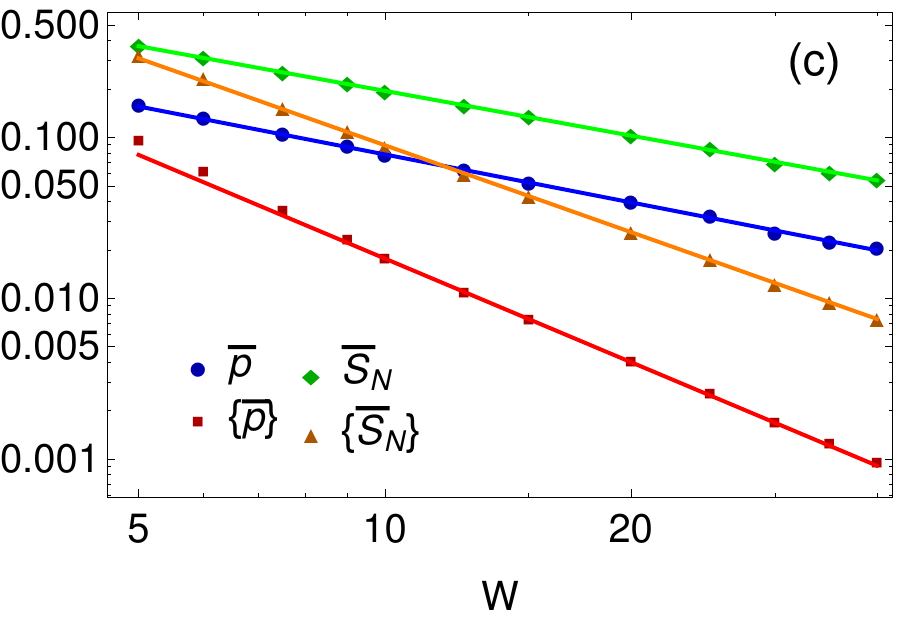}
\caption{Plot showing various fits of steady state quantities. (a) For $L=12$ the plot shows the comparison of scaling laws for the various measures of steady state values considered in the main text in Sec. \ref{steady}.  The solid lines are fits of the form $a_1-a_2/W^b$. (b) Plots showing how $b$ changes with system size $L$. Fits are of the form $c_1-c_2/L$. See text for $c_1$ and $c_2$ in each case. (c) Similar fits to those in panel (a) but with only $q=1$ initial states. See text for details. }
\label{figappb1}
\end{figure*}
In Fig. \ref{figappb1}(a) we show a comparison of scaling laws of various statistical quantities we have used to quantify the steady state behaviour of the system in Sec. \ref{steady}. $\bar{p}$ denotes mean steady state probability of subsystem $A$ having different number of particles than the initial state, $\bar{S}_N$ denotes mean steady state $\SN$ and $\tilde{S}_N$ denotes mean long time average of $\SN$ . These three quantities show extremely similar power law decay with $W$. We fit the quantities to a general form $a_1-a_2/W^b$, where $b$ is given by $1.1, 1.07, 1.13$ for $\bar{p}$, $\bar{S}_N$ and $\tilde{S}_N$  respectively . The median values denoted by $\{\bar{p}\}$, $\{\tilde{S}_N\}$ and  $\{\bar{S}_N\}$ again have similar behaviour, but they are not as close to each other as the mean. $\{\bar{p}\}$ shows a faster decay with  $b= 2.6$ while $\{\bar{S}_N\}$ and  $\{\tilde{S}_N\}$ has slightly slower decay with $b= 2.2$ and $2.38$ respectively. \\

In Fig.~\ref{figappb1}(b) we show how $b$ grows with system size for four selected quantities from Fig. \ref{figappb1}(a). With the power law fit of the quantities as described, $b$ shows a very slow growth with system size. The growth of $b$ can be fitted into the form $c_1-c_2/L$ where $c_1$ denotes the extrapolated value of $b$ at $L\rightarrow \infty$. The values of $c_1$ are $2.34$,$1.13$,$2.74$ and $1.17$ for $\{\bar{S}_N\}$, $\bar{S}_N$, $\{\bar{p}\}$ and $\bar{p}$ respectively. \\

The slight rise in values of $b$ with increasing $L$ can be attributed to the choice of initial states. We discussed in Sec. \ref{short} , how different initial states can be grouped via a quantity we denote as $q$. We have also elaborated on how states with different $q$ generally show markedly different number entropy values at all timescales. Consequently, pooling the different states together during steady state calculations has an effect on the quantities we are interested in. For example, in the main text we showed how one can get a $1/W^2$ decay of $\{\bar{p}\}$ from a two-state model. But in Fig. \ref{figappb1}(a) we see a discrepancy of $0.6$ with this result. However in Fig. \ref{figappb1}(c) we plot the mean and median of the respective steady state number entropy and probability distributions for a system of size $L=12$ when we just consider $q=1$ states. The fit value of $b$ here are $0.99$, $2.14$, $0.92$ and $1.79$ for $\bar{p}$, $\{\bar{p}\}$, $\bar{S}_N$ and $\{\bar{S}_N\}$ respectively. We clearly see the slope  of $\{\bar{p}\}$ vs $W$ to show a value closer to $2$ than that in Fig. \ref{figappb1}(a). In fact one can show the change in $b$ with system size is negligible if one takes into account only $q=1$ states.\\

Another takeaway from these plots is that the $\bar{S}_N$ and $\{ \bar{S}_N \}$ always shows a power law decrease with $W$ with a slightly lower value of $b$ than $\bar{p}$ and $\{\bar{p}\}$. This is also to be expected because $\SN$ is actually a non-linear function of $p$, and a small deviation is possible even with a skewed distribution of $p$. The fact that the differences in  values of $b$ are very small between these two quantities validates the linear approximation used in the main text.

\section{Filtering procedure and identification of timescales}
\label{appC}
 \begin{figure}
 \centering{
\vspace{0.1in} \includegraphics[width=0.8 \columnwidth]{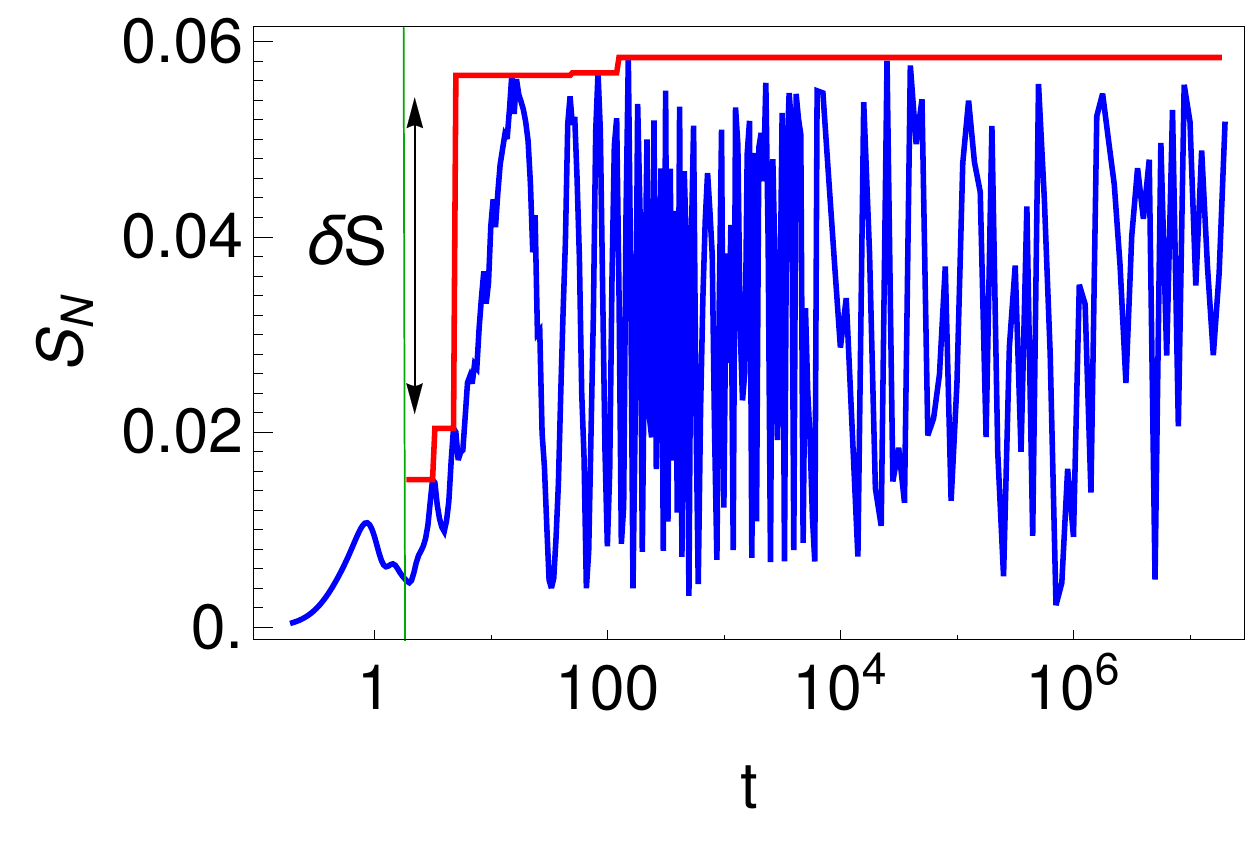}}
 \caption{ Figure showing the approximation technique used to determine the position and height $\delta S$ of a jump using representative data from a randomly selected configuration of disorder and initial state. }
 \label{fig3a}
 \end{figure}
 
  \begin{figure}
 \centering{
 \includegraphics[width=0.8 \columnwidth]{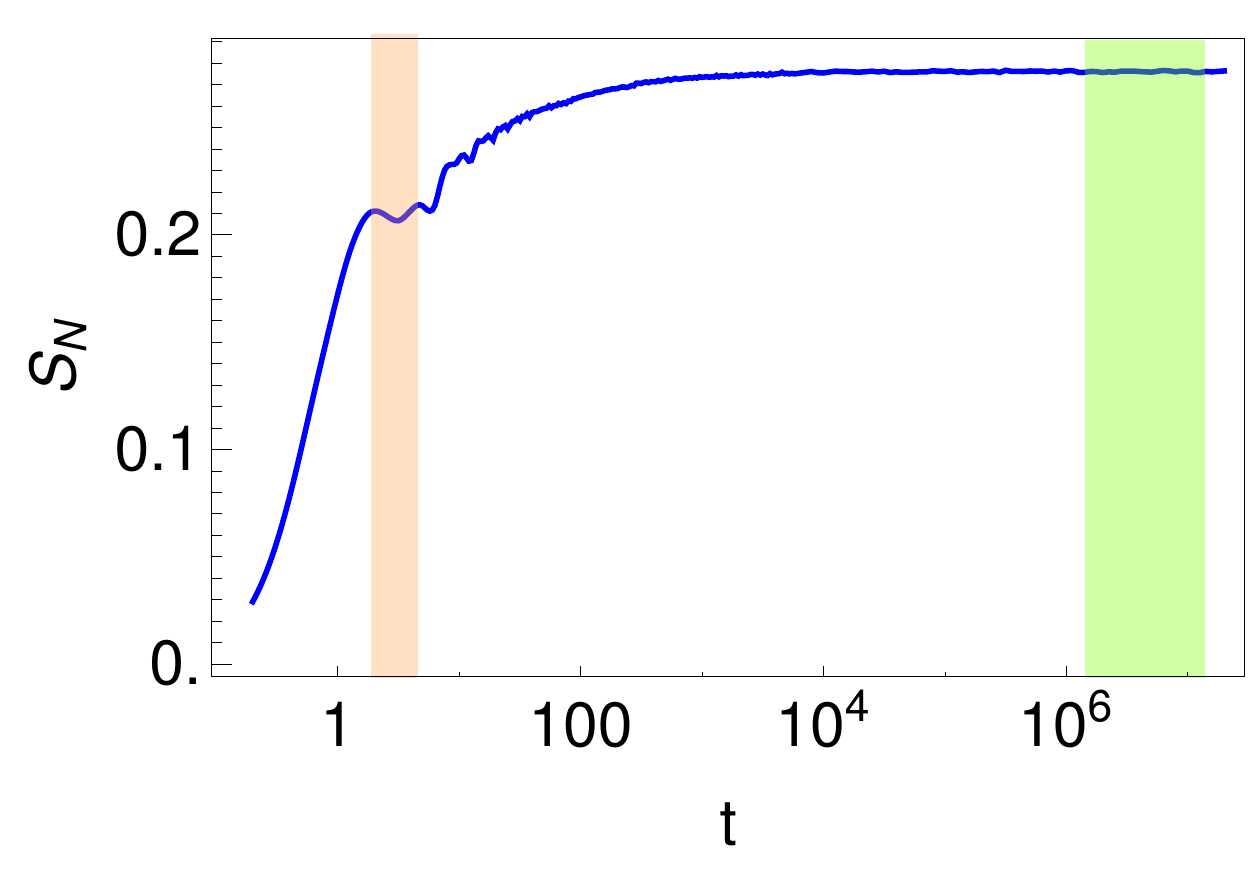}}
 \caption{ Plot of disorder-averaged $\SN(t)$ for a representative data with $L=12$ and $W=5$ showing the region we identify as short-time regime shaded in orange,  and the region we identify as long-time regime shaded in green, used while calculating $\Delta S$ in Sec. \ref{inter}. }
 \label{fig3b}
 \end{figure}
In  Fig. \ref{fig3a} we show how we approximate the data for $\SN$ for individual configurations to quantify the jumps. We first smoothen the plot by considering an envelope of the oscillatory function, which is created by using the maximum value of $\SN(t)$ until the $t$ under consideration. This approximates $\SN$ via a series of step functions as shown in the plot with the red line. We then calculate the difference between successive plateaus of this envelope and denote the differences by $\delta S$; we then look for cases where $\delta S>0.1$. The green line denotes the time $t=2$, which is approximately the time from which we start the filtering process. This is done as we intend to ignore the short-time rise of $\SN$ caused due to the nearest neighbour resonances during the analysis in Sec. \ref{inter}. \\

In the Fig. \ref{fig3b} we pictorially show the two timescales involved in calculating the total increase in mean $\SN$, denoted by $\Delta S$. Initially due to Rabi oscillations $\SN$ is oscillatory and we quantify the short time value of $\SN$ (which we roughly assume occurring when one particle has crossed the subsystem boundary), by taking the mean $\SN$ between the peaks of the first two oscillations. This is denoted by the orange shaded region. For long-time data we average between $t \sim 10^6-10^7$ where there is almost no change in $\SN$. This is denoted by the green shaded region\\

\section{Can one get the growth of $\SN$ from LIOM picture?}
\label{LIOM}
While in this work we have studied the isotropic Heisenberg model, one can approach this problem in a model-independent manner, using the LIOMs picture of the MBL phase.~\cite{PhysRevB.90.174202,PhysRevLett.111.127201,serbyn14,anto15} However unlike the study of $S$ via the dephasing model, one needs to rotate back from the LIOMs basis to the computational basis to extract the number entropy from the model. 

One way to approach this is to use a quantum circuit model. One can start with a layer of two site unitary quantum gates, conserving particle number, parameterized by an angle $\phi$, and add several such layers to generate a unitary matrix $U$ replacing the eigenvectors which form the rotation matrix connecting the two bases. In principle, the parameter $\phi$ is free, but to simulate locality, one should use a small value of $\phi=\phi_1$ at the first layer and then use $\phi_l=\phi_1 e^{-l/\zeta}$ for the subsequent layers ($\zeta$ is the localization length taken from the LIOM model), where $l$ is used to denote the layer index. Tuning $\phi_1$ and the maximum allowed value of $l$, it might seem that localized systems of interest can be approximated.

 However we found that even for low $W$ (but within MBL phase) one cannot obtain the correct late-time growth of $\SN$ reported. One does see a small growth in $\SN$ at low $W$ for this setup but it is driven by movement of particles within the localization length. It is important to note that in such a setup, the approximated eigenvectors are forcibly made to show generic exponential decay with hamming distance, and do not include the special resonant eigenvectors. This causes the saturation of $\SN$ to occur earlier than what is seen from the numerics, since the increase at later times driven by resonances are not captured by this setup. Upon increasing $W$ the small increase  gradually vanishes, as localization length becomes smaller and there are no resonances to drive the growth of $\SN$. Hence we note that the LIOMs model with the aid of a generic quantum circuit cannot easily replicate the results discussed in this work.
 
\section{Comparison with Anderson localization}
\label{anderson}
In this appendix we will show results of $\SN$ obtained from non-interacting model which shows Anderson localization in 1D. It is known that there is no particle transport at distances much larger than the localization length $\zeta$ in this system.  We shall find that the steady state mean and median $\SN$ behave quite similarly to the interacting case. However we will see that the number of resonant disorder cases is significantly lower here, especially for long-range resonances.\\
\begin{figure}
\includegraphics[width=0.7 \columnwidth]{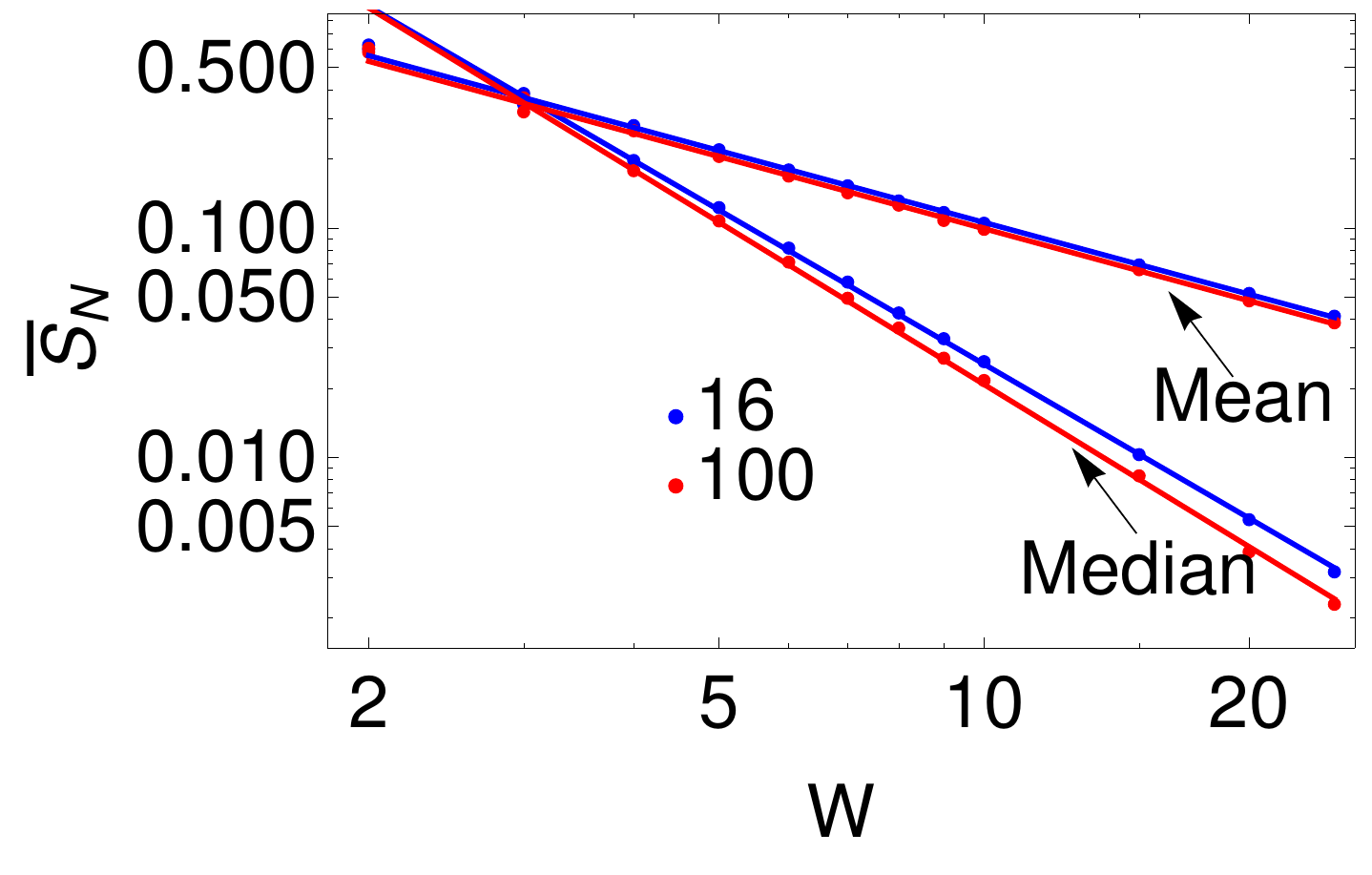}\\
\caption{The plot shows the statistics of mean and median of $\bar{S}_N$ for an Anderson model with system size $L=16$ and $L=100$. The lines indicate fits to power laws. See text for exponents.}
\label{andsteady}
\end{figure}
\begin{figure}
\hspace{-0.65 in}\includegraphics[width=0.55 \columnwidth]{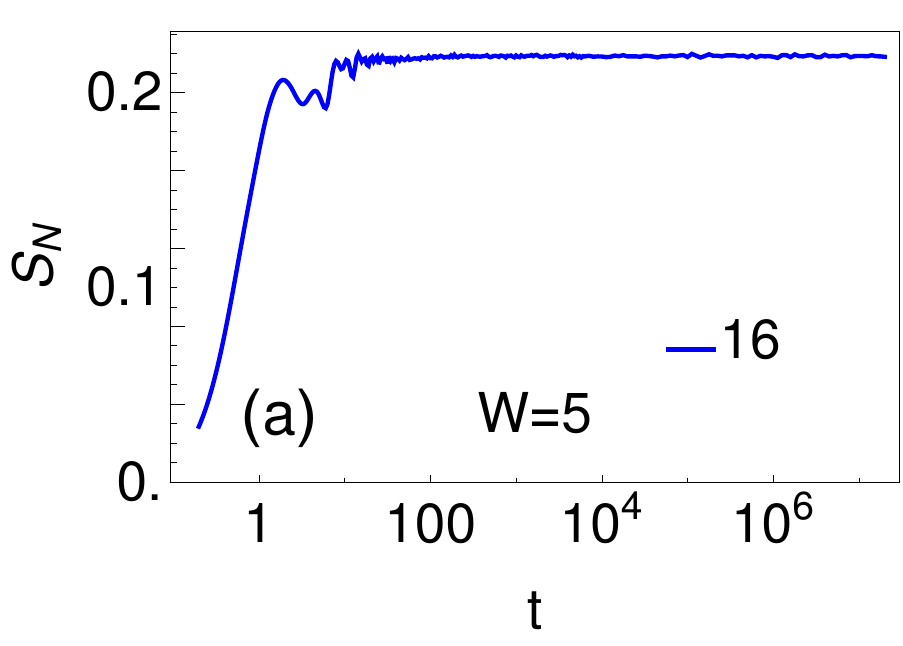}\\
\includegraphics[width=0.80 \columnwidth]{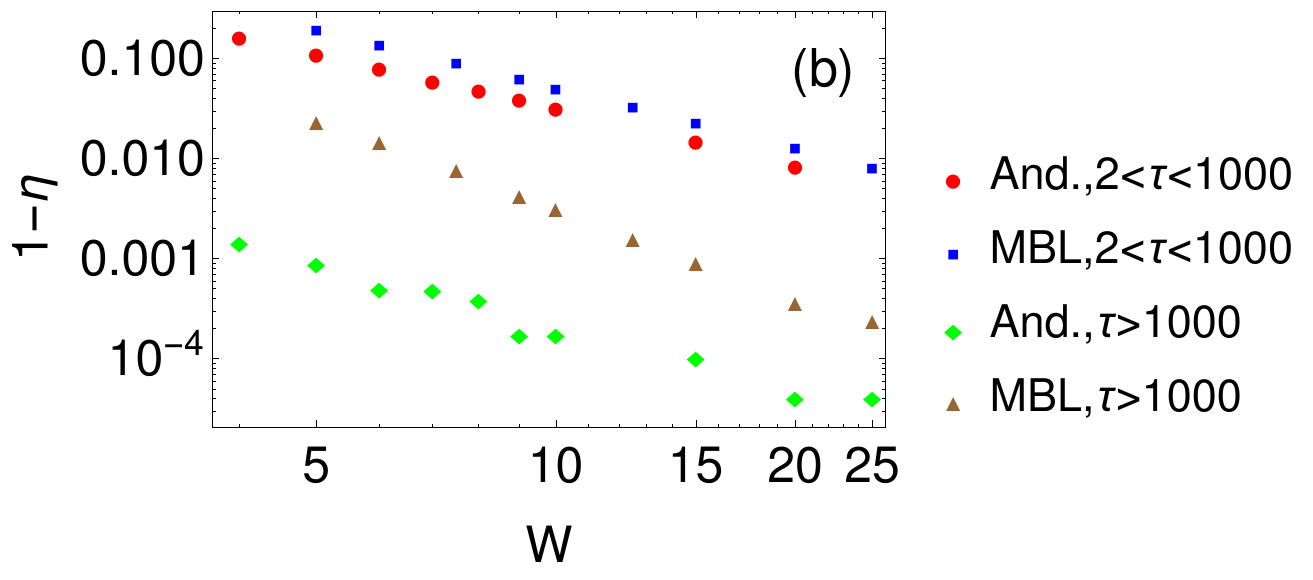}
\caption{Plots showing results for Anderson model for $L=16$. (a) Plot showing disorder averaged $\SN$ vs $t$ for an Anderson-localized system. (b) Plot showing comparison of fraction ($1-\eta$) of configurations showing $\delta S>0.1$ jumps for Anderson and MBL models at two different timescales ($\tau$).}
\label{andfig}
\end{figure}
In Fig. \ref{andsteady}, we plot the mean and median of the steady state number entropy,  $\bar{S}_N$ for the Anderson localized system at two different system sizes. For size $16$, the fit denoted by blue lines are $W^{-1.04}$ for mean and $W^{-2.23}$ for median. For size $100$, fits denoted by red lines are $W^{-1.04}$ and $W^{-2.34}$ for mean and median respectively. These fits are extremely close to what we have seen for the interacting case, which further affirms our statement that there is no particle transport over long distances in the interacting case as well. \\

In Fig. \ref{andfig}(a), we first show how in an Anderson localized system, the slow growth in $\SN$ is not visible. Then in Fig. \ref{andfig}(b) we compare the ratio of number of resonant disorder configurations to the total number of configurations in the Anderson localized case and the MBL case. We find that, as expected, the MBL system shows a higher number of resonances due to higher connectivity of Hilbert space. The more interesting feature however is in the Anderson-localized case; long-range resonances which drive the growth in $\SN$ at later times are significantly lower in number. The possible cause for this is the different selection rules in the Hamiltonian for interacting and non-interacting systems. Denoting approximate time of the jump by $\tau$ we quantify this by calculating the fraction of total cases that are resonant cases at $\tau>1000$. We see that this number is heavily suppressed for the Anderson localized case compared to the MBL case. For the timescales less than $1000$ the difference is not that large between the two. This explains why one sees a much smaller growth of $\SN$ in the Anderson localized case at longer timescales compared to MBL. Note that the number of long time resonant configurations at $W=4$ for the Anderson case is almost the same as the number of cases in $W=15$ in the MBL case, where we know that the rise of $\SN$ is very small (See Fig. \ref{fig1} of the main text).
\end{document}